\documentclass[aps,pre,twocolumn,groupedaddress,floatfix]{revtex4-1}

\usepackage{chemfig}
\usepackage{amsmath}
\usepackage{amsfonts}
\usepackage{graphicx}
\usepackage{epstopdf}	

\newcommand{\free}{F}
\newcommand{\field}{E}
\newcommand{\energy}{\epsilon}
\newcommand{\param}{\mu}

\newcommand{\control}{\lambda}

\usepackage{graphicx}
\usepackage{amssymb}
\usepackage{epstopdf}
\begin{document}


\title{Far-from-Equilibrium Distribution from Near-Steady-State Work Fluctuations}


\author{Robert Marsland III}
\author{Jeremy England}
\affiliation{Physics of Living Systems Group, Massachusetts Institute of Technology, 400 Technology Square, Cambridge, MA 02139}


\date{\today}

\begin{abstract}
A longstanding goal of nonequilibrium statistical mechanics has been to extend the conceptual power of the Boltzmann distribution to driven systems. We report some new progress towards this goal. Instead of writing the nonequilibrium steady-state distribution in terms of perturbations around thermal equilibrium, we start from the linearized driven dynamics of observables about their stable fixed point, and expand in the strength of the nonlinearities encountered during typical fluctuations away from the fixed point. The first terms in this expansion retain the simplicity of known expansions about equilibrium, but can correctly describe the statistics of a certain class of systems even under strong driving. We illustrate this approach by comparison with a  numerical simulation of a sheared Brownian colloid, where we find that the first two terms in our expansion are sufficient to account for the shear thinning behavior at high shear rates.
\end{abstract}




\maketitle
\section{Introduction}
For over a century, the formalism of equilibrium statistical mechanics has provided a powerful means to explain how the macroscopic properties of many-body systems at thermal equilibrium arise from the microscopic interactions that occur among their constituent parts.  The centerpiece in this approach is the Boltzmann distribution, which posits that the probability of observing an equilibrated system with energy $\energy$ in microstate $x$ at temperature $k_B T=1/\beta$ is proportional to the so-called ``Boltzmann weight" $p_{\textnormal{bz}}(x)\propto \exp[-\beta \energy (x)]$.  The key assumption used in deriving the Boltzmann distribution is that the system has spent an ``ergodically" long time in contact with its surrounding heat bath, so that the combined set-up of bath and system is equally likely to be in any arrangement that is allowed by conservation of energy.  As a result, a quantity evaluated for the system at one instant in time (namely, $\energy (x)$) can immediately be translated into a probability of occurrence for the state $x$. This microscopic result can be coarse-grained to yield the probability of observing the system in a given macroscopic state $X$, defined as a set of microstates that share the same values of some observable properties. The coarse-grained probability $p_{\textnormal{bz}}(X)\propto e^{-\beta \free (X)}$ can then be written in terms of a free energy $\free (X)=-k_BT \ln\left(\sum_{x\in X} e^{-\beta \energy (x)}\right)$.

Once time-varying fields drive the system from equilibrium by changing the energies $\energy (x,t)$ on timescales comparable to the system's relaxation time, the story must necessarily become more complicated.  In the arbitrary nonequilibrium scenario, the probability of being at a given location in phase space at time $t$ clearly can depend strongly on where the system was at some earlier moment.  There is, however, a tempting special case to consider even when the Boltzmann distribution does not apply: in circumstances where $\energy (x,t)$ is periodic, the system may still ergodically lose its memory of initial conditions after enough time in contact with the fluctuating bath.  In such a case, it is reasonable to consider whether the Boltzmann distribution admits a generalization, in which the probability of observing the system in a particular state after the memory of the initial state is lost can still be related exactly to some function of thermodynamic observables. 

Yamada and Kawasaki answered this question in the affirmative almost fifty years ago, when they derived an effective partition function for a generic nonequilibrium steady state in terms of correlations in the currents of conserved quantities passing through the system \cite{Yamada1967}, launching a fruitful field of research on the exact microscopic distribution in driven steady states \cite{Morriss1985, Morriss1988, Evans2008, Crooks1999}. It is now well known that the simplicity of the Boltzmann weight cannot be reproduced in the microscopic probability distribution for an arbitrary driven steady state, which depends in general on all orders of the time correlations in the currents over the system's past history.

Thus any attempt to uncover simple principles beneath the statistics of nonequilibrium steady states must begin by specifying a particular regime of applicability, where certain simplifying approximations become valid. The near-equilibrium regime was the first to receive careful study, leading to an elegant representation of the steady-state distribution in terms of the dissipation due to externally imposed thermal gradients, chemical potential gradients, and velocity fields \cite{McLennan1960}. This ``McLennan ensemble'' applies to a wide range of near-equilibrium systems, and can be obtained through a variety of independent routes (cf. \cite{Lebowitz1960}). Most recently, it has been shown that this form can be derived in an especially transparent way from the assumption of ``microscopic reversibility,'' which holds for a wide class of physical systems \cite{Crooks1999,Komatsu2009,Maes2010}. 


When external drives become arbitrarily strong, the steady-state distribution of observables in a generic physical system no longer follows this form. But some intution about this regime can be built up around a special case, where a simple expression for the distribution does exist for arbitrarily strong driving, with the general expression written in terms of perturbations about this case. In this paper, we derive such an expansion about the case where fluctuations in the instantaneous dissipative current in the driven system obey a linear overdamped Langevin equation with additive white noise. In this scenario, a form essentially equivalent to the McLennan ensemble can be shown to hold regardless of the drive strength, and to be compatible with large departures from the predictions of linear-response theory.

This approach bears some resemblances to macroscopic fluctuation theory, which derives various statistical properties of far-from-equilibrium macroscopic systems from some minimal restrictions on the form of the macroscopic dynamics and the character of the current fluctuations \cite{Bertini2015}. In particular, a simple form for the driven steady-state distribution is obtained that looks very similar to the McLennan ensemble, but holds for arbitrarily strong driving with arbitrary nonlinearities \cite{Bertini2012}. This result is obtained through a decomposition of the dissipative current into ``symmetric'' and ``asymmetric'' parts, which purchases a broader range of applicability by sacrificing the immediate physical meaning of the terms in the original McLennan form. We do not take advantage of this decomposition, but instead seek a representation of the distribution entirely in terms of the equilibrium free energy and the statistics of the bare externally applied work.

We start in section \ref{sec:setup} by deriving an exact expression for the steady-state distribution of an arbitrary observable in a system with microscopic reversibility whose distribution relaxes exponentially to a unique stationary state. Then in section \ref{sec:general} we evaluate this expression in our special case, and compute the first correction in the perturbation expansion. In sections \ref{sec:system} and \ref{sec:work} we apply this expansion to a sheared colloid, as a specific example of a strongly driven system. Section \ref{sec:results} gives a quantitative comparison with a numerical simulation of the sheared colloid, showing that the McLennan-like part is sufficient to reproduce the qualitative non-linear response behavior, and that the first correction term gives good quantitative agreement deep into the shear thinning regime. Finally, in section \ref{sec:conclusion} we describe the thermodynamic intuition that can be extracted from this new form for the distribution, and define some avenues for further investigation.

\section{Derivation of Distribution}
\label{sec:setup}
\subsection{Microscopic Reversibility}
Consider a generic physical system coupled to a large ``heat bath''
of inverse temperature $\beta=1/k_BT$, but otherwise isolated, so that the total energy of the system plus bath can only be changed by external manipulation of a specified set of ``control parameters'' $\control$. These control parameters directly affect only the system energy $\energy (x,\control)$ and not the bath. As usually assumed in statistical mechanics, the potential energy of the interactions of system components with bath components is taken to be small compared with the energy in the system, so that the total energy can be cleanly divided between the energy ``in the system'' and the energy ``in the bath.'' The bath volume is also held fixed, and the system volume is only allowed to vary if it is chosen as one of the control parameters $\control$. 

The microstate $x$ of the system evolves according to a stochastic process, due to its interactions with the fluctuating heat bath. If the combined setup is modeled with classical mechanics, $x$ specifies generalized positions and momenta of all degrees of freedom in the system (not including the bath degrees of freedom). $x$ can also be taken to be a discrete microstate label, with jumps between microstates governed by a rate matrix $W_{xx'}(\control(t))$. In this case, thermodynamic consistency is imposed by requiring that the rate matrix should eventually equilibrate the system to the Boltzmann distribution if the $\control$'s are held fixed. 

We keep the system out of equilibrium by varying the $\control$'s according to a given protocol, which gives the microstate energies $\energy (x,\control(t))$ an explicit time dependence. When we average over trajectories below, we will always assume that the $\control(t)$ protocol is held fixed. 

Building on the work of C. Jarzynski \cite{Jarzynski1997}, G. Crooks has shown that a number of important results concerning the nonequilibrium behavior of such a system can be derived from what he calls the
``microscopic reversibility'' condition \cite{Crooks1999}:
\begin{align}
\label{crooks1}
\frac{p_R[x^*(\Delta t-t)|x^*_2]}{p_F[x(t)|x_1]}=e^{-\beta Q_F[x(t)]}.
\end{align}
Here $x(t)$ is a system trajectory of duration $\Delta t$ and the heat $Q_F[x(t)]$ is the energy transferred from the system to the bath over the course of that trajectory.
The left-hand side contains the
probability of taking the time-reversed path $x^*(\Delta t-t)$ given a
starting state $x^*_2$, divided by the probability of taking the forward
path $x(t)$ given the starting state $x_1$. The $*$ indicates
time-reversal of the microstate (changing the signs of all the momenta in the classical model). The $R$ and
$F$ subscripts refer to the driving protocol $\control(t)$: the $F$ probabilities and heat are computed with the
protocol forward from time $-\Delta t$ to time 0, and the $R$ probabilities
have it running in reverse from time 0 to time $-\Delta t$. In the sheared colloid we will analyze below, the $R$ and $F$ quantities are computed
with the shear applied in opposite directions.

Crooks showed that condition (\ref{crooks1}) holds for stochastic dynamics of discrete states under the thermodynamic consistency requirement given above \cite{Crooks1999}. This relation can also be derived directly from the time-reversibility and phase space conservation of Hamiltonian dynamics, if the combined system-plus-bath setup can be treated as a closed Hamiltonian system driven by an explicit time-dependence in the system Hamiltonian (cf. \cite{Jarzynski2000} for the basic approach, although a slightly different result is discussed there). The expression can be generalized to allow particle fluxes into and out of chemical baths, in which case an extra term involving chemical potentials must be added to the $Q_F$ in the exponent \cite{Qian20052,Qian2013,Gaspard2004}.

\subsection{Coarse Graining}

Now we group the system microstates $x$ according to some observable properties, following the approach detailed in \cite{England2013}. We can give each group a unique name, which will be generically represented by the capital letter $X$. Below, in the analysis of colloidal steady states, $X$ will stand for a single number (the mean shear stress at the wall) characterizing the configuration of all the colloidal particles. In general, $X$ represents a label for some group of microstates, which could be picked out using any definite procedure. $X^*$ will refer to the group consisting of the time-reversed versions $x^*$ of all the microstates in $X$. 

The coarse-grained version of equation (\ref{crooks1}) involves transition probabilities among the different groups labeled by different $X$ values. This probability will in general depend on how the initial state was prepared, since different protocols will give rise to different probability distributions of microstates within the macrostate. In this paper, however, we consider only systems with finite relaxation times, and our goal is to analyze the steady state that is reached after all correlations between current $X$ values and the initial conditions have died out. In such a case, the choice of initial distribution becomes irrelevant to the steady-state statistics, and we can choose the distribution that gives the most useful form for the final steady-state distribution of $X$.

We will call the initial distributions for the forward and reverse trajectories $p_1(x)$ and $p_2(x)$ respectively, and choose them to be Boltzmann distributions $p(x) =\exp[-\beta (\energy (x,\control(t)) - \free (X,\control(t)))]$ over the microstates in the respective macrostates $X_1$ and $X_2^*$. $\free (X,\control(t)) = -k_BT \ln \sum_{x\in X} \exp[-\beta \energy (x,\control(t))]$ gives the proper normalization for a distribution defined only over microstates $x$ in macrostate $X$ (where the sum becomes an integral in the classical case). To investigate steady-state behavior, we must vary these fields periodically, and consider trajectories whose duration $\Delta t$ is an integer multiple of the period. This implies that $\energy (x,\control(-\Delta t)) = \energy (x,\control(0))$ and $\free (x,\control(-\Delta t)) = \free (x,\control(0))$, so for both cases we can drop the time-dependence from the notation.

With these distributions in hand, we multiply equation (\ref{crooks1}) by the denominator of the left-hand side and by $p_2(x_2^*)$, then integrate over all trajectories $x(t)$ connecting states $x_1$ in a given macrostate $X_1$ to states $x_2$ in another macrostate $X_2$:
\begin{align}
\label{paths}
&\int_{X_1\to X_2} \mathcal{D}[x(t)] p_2(x^*_2)p_R[x^*(\Delta t-t)|x^*_2]\nonumber\\
&=\int_{X_1\to X_2} \mathcal{D}[x(t)] e^{-\beta Q_F[x(t)]}\frac{p_2(x^*_2)}{p_1(x_1)}p_1(x_1)p_F[x(t)|x_1].
\end{align}
We can simplify this expression by introducing the macroscopic transition probabilities 
\begin{align}
\label{macprobs}
\pi_F (X_1\to X_2) &\equiv \int_{X_1\to X_2}  \mathcal{D}[x(t)]p_1(x_1)p_F[x(t)|x_1]\\
\pi_R(X_2^*\to X_1^*) &\equiv \int_{X_1\to X_2} \mathcal{D}[x(t)]p_2(x_2^*)p_R[x^*(\Delta t-t)|x^*_2]
\end{align}
which are defined as the sums of the probabilities of all microtrajectories $x(t)$ that accomplish the indicated macroscopic transition, given that the system begins in the indicated probability distribution over microstates ($x_1$ or $x_2^*$) in the starting macrostate ($X_1$ or $X_2^*$). Using these definitions to normalize the distributions over trajectories in the integrands in equation (\ref{paths}), we find:
\begin{align}
\pi_R(X_2\to X_1)=&\langle e^{-\beta Q_F[x(t)]+\ln \frac{p_2(x^*_2)}{p_1(x_1)}}\rangle_{X_1\to X_2}\nonumber\\
&\times \pi_F(X_1\to X_2).
\end{align}
The average $\langle \cdot\rangle_{X_1\to X_2}$ is over all trajectories $x(t)$ connecting some microstate $x_1$ in $X_1$ (chosen from distribution $p_1$) to some other microstate $x_2$ in $X_2$. 

Now we insert the explicit expressions for $p_1$ and $p_2$, to find
\begin{align}
\frac{\pi_R(X_2^*\to X_1^*)}{\pi_F(X_1\to X_2)}=& e^{\beta[\free (X_2)-\free (X_1)]}\nonumber \\
&\times\langle e^{-\beta [Q_F[x(t)]-\energy (x_1)+\energy (x_2)]}\rangle_{X_1\to X_2}.
\end{align}
We have dropped the $*$ in the arguments of $\energy$ and $\free$, because the energy is symmetric under reversal of the signs of the momentum coordinates. Now we note that by conservation of energy, the work done on the system by the variation of the control parameters $\control(t)$ over a trajectory $x(t)$ from $x_1$ to $x_2$ satisfies $W_F = Q_F[x(t)]+\energy (x_2)-\energy (x_1)$. We can thus rewrite the above expression as
\begin{align}
\frac{\pi_R(X_2^*\to X_1^*)}{\pi_F(X_1\to X_2)}=&e^{\beta[\free (X_2)-\free (X_1)]}\nonumber \\
&\times\langle e^{-\beta W_F[x(t)]}\rangle_{X_1\to X_2}.
\label{crooksmacro}
\end{align}
The motivation for our choice of the Boltzmann distribution for $p_1$ and $p_2$ is precisely because it replaces the heat in the exponent with the work. Since the work is zero in the undriven case, where the control parameters are fixed, this choice splits the right-hand side cleanly into two factors, an equilibrium contribution and a nonequilibrium correction. Other choices could be made for these distributions, and they would generate valid alternative forms of the steady-state distribution.

We can use this expression to compare the probabilities of forward transitions from one state $X_0$ to two different states $X_1$ and $X_2$
\begin{align}
\ln \frac{\pi_F(X_0\to X_1)}{\pi_F(X_0\to X_2)}=&\beta [\free (X_2)-\free (X_1)]\nonumber\\
&-\ln \frac{ \langle e^{-\beta W_F[x(t)]}\rangle_{X_0\to X_1}}{ \langle e^{-\beta W_F[x(t)]}\rangle_{X_0\to X_2}}\nonumber\\
&+\ln \frac{\pi_R(X_1^*\to X_0^*)}{\pi_R(X_2^*\to X_0^*)}.
\label{macrotrans}
\end{align}
This expression is interesting in its own right, showing how the relative probabilities of different future possibilities depend not only on the free energies of the possible future states, but also on the work done on the way there and on the ``durability'' measure contained in the reverse probabilities (see \cite{England2014} for a detailed analysis of the physical implications of the corresponding terms in a closely related expression). We now specialize to a system in which temporal correlations decay exponentially (or faster) with a finite relaxation time $\tau$. In this case, $\pi(X_i\to X_f)$ must become independent of $X_i$ for $\Delta t\gg \tau$, and simply be equal to the steady-state probability $p_{\rm ss}(X_f)$ of being in the final state:
\begin{align}
\ln \frac{p^F_{\rm ss}(X_1)}{p^F_{\rm ss}(X_2)}=&\beta [\free (X_2)-\free (X_1)]\nonumber\\
&-\lim_{\Delta t\to \infty}\ln \frac{ \langle e^{-\beta W_F[x(t)]}\rangle_{X_0\to X_1}}{ \langle e^{-\beta W_F[x(t)]}\rangle_{X_0\to X_2}}\nonumber\\
&+\ln \frac{p^R_{\rm ss}( X_0^*)}{p^R_{\rm ss}(X_0^*)}.
\end{align}
From this equation we can directly extract $p^F_{\rm ss}(X)$ up to an overall constant $\mathcal{N}$ that is independent of $X$:
\begin{align}
\ln p_{\rm ss}(X)=&-\beta \free (X)-\lim_{\Delta t\to \infty}\ln \langle e^{-\beta W}\rangle_{X_0\to X}+\mathcal{N}
\end{align}
where we have dropped the $F$'s from $p_{\rm ss}$ and $W$ because we will only be considering ``forward'' quantities from now on.

Even though the work $W$ in the exponent increases without bound as $\Delta t\to \infty$, this expression is well-defined if one takes care to perform the average before taking the limit. For exponentially relaxing systems, $\langle e^{-\beta W} \rangle_{X_0 \to X}$ converges to a finite value as $\Delta t$ increases beyond the relaxation time $\tau$. This can be seen by considering the behavior of the LHS of equation (\ref{crooksmacro}) under these conditions: $\pi_F(X_1 \to X_2)$ and $\pi_R(X_2^* \to X_1^*)$ approach the finite limiting values of $p^F_{\rm ss}(X_2)$ and $p^R_{\rm ss}(X_1^*)$, respectively, for $\Delta t \gg \tau$.

The individual cumulants of $W$ do in general become infinite as $\Delta t\to \infty$, however. To obtain a useful series representation of the exponential average term, we can add the constant, finite quantity $\lim_{\Delta t\to \infty} \ln \langle e^{-\beta W}\rangle_{X_0\to X_0}$ to the RHS, and compensate by adjusting the normalization constant $\mathcal{N}$. Representing both exponential average terms by their cumulant expansions, we obtain:

\begin{align}
\label{pss0}
\ln p_{\rm ss}(X)=&-\beta \free (X) -\lim_{\Delta t\to \infty}\bigg( \sum_{n=1}^\infty \frac{(-\beta)^n}{n!}\langle W^n\rangle^c_{X_0\to X}\nonumber\\
&- \sum_{n=1}^\infty \frac{(-\beta)^n}{n!}\langle W^n\rangle^c_{X_0\to X_0}\bigg)+\mathcal{N}. 
\end{align}

We can now rewrite this expression in terms of the finite differences between the cumulants for the trajectories ending in $X$ and the corresponding cumulants for the trajectories ending in $X_0$:
\begin{align}
\label{renormalized}
\Delta \langle W^n\rangle^c (X) &\equiv \lim_{\Delta t\to \infty}\left( \langle W^n\rangle_{X_0\to X}^c-\langle W^n\rangle^c_{X_0\to X_0}\right) \nonumber\\
&=\lim_{\Delta t\to \infty}\left( \langle W^n\rangle_{{\rm ss}\to X}^c-\langle W^n\rangle^c_{{\rm ss}\to X_0}\right).
\end{align}
Since we have assumed that the system's state becomes decorrelated from its past history after a finite time $\tau$, the expression in parentheses becomes independent of $\Delta t$ and remains finite as we take $\Delta t\to\infty$. In the second line, the ${\rm ss}\to X$ averages are over the trajectories whose initial conditions are sampled form the steady state distribution, and that end in state $X$. The equality follows from the fact that the contribution of the initial relaxation of the system from $X_0$ to the steady state is the same for both terms on the RHS of the first line.

Thus we obtain:
\begin{align}
\label{pss}
\ln p_{\rm ss}(X)=&-\beta \free (X)-\sum_{n=1}^\infty \frac{(-\beta)^n}{n!}\Delta \langle W^n\rangle^c  (X)+\mathcal{N}. 
\end{align}
Note that if $X$ is an observable quantified by a continuous parameter, $p_{\rm ss}(X)$ can be regarded as a probability density function for $X$. In this case, the LHS should strictly be written as $\ln [p_{\rm ss}(X)\delta X]$, where a microstate counts as part of macrostate $X$ if the value of the observable lies within $\delta X$ of $X$. But the $\delta X$ can be chosen to be the same for all $X$, and rolled into the overall constant $\mathcal{N}$. 

Aside from the coarse-graining step, the general procedure we have followed thus far for extracting a steady-state distribution from microscopic reversibility and expanding in cumulants resembles the derivation by Komatsu \emph{et. al} of an expansion about equilibrium in the strength of the driving field \cite{Komatsu2009}. Expressions like (\ref{pss}) obtained in this way contain cumulants of all orders, and only provide new physical insight if the series converges rapidly. Komatsu \emph{et al.} show how a different way of writing the microscopic reversibility assumption leads to an expansion about equilibrium that converges particularly rapidly, and is accurate to second order in the strength of the drive without including any cumulants of higher order than 1 \cite{Komatsu2009}. We are taking a different approach, treating equation (\ref{pss}) as an expansion in the size of the nonlinearities in the coarse-grained equations of motion near the steady state. The coarse-graining allows the parameters of the linearized dynamics to have a non-trivial dependence on the strength of the drive, so that the predictions of near-equilibrium linear response theory can break down while our expansion parameter is still near zero. In the next section, we lay out the details of our proposed expansion in terms of a Langevin model for the coarse-grained dynamics. Then we will use a simulated sheared colloid to illustrate a concrete case where the new expansion converges rapidly in a strongly driven system. 

Writing the distribution in terms of the $\Delta \langle W^n\rangle^c$'s is convenient for the initial presentation of the theory, but for comparison with the colloid simulation, we need to work with finite $\Delta t$'s. Thus we define
\begin{align}
  \label{sswork}
  \langle W^n\rangle^c_{\Delta t}(X) &\equiv \langle W^n\rangle^c_{{\rm ss}\to X} \\
&\approx \Delta \langle W^n\rangle^c(X) + \langle W^n\rangle^c_{{\rm ss}\to X_0}\nonumber
\end{align}
where the averages are all over trajectories of length $\Delta t$, and the approximation holds for $\Delta t\gg \tau$. Since the second term in the second line is independent of $X$, replacing $\Delta \langle W^n\rangle^c(X)$ with $\langle W^n\rangle^c_{\Delta t}(X)$ in any expression for the steady-state distribution only affects the normalization. In terms of this new quantity, we thus obtain the alternative form
\begin{align}
\label{pssemp}
\ln p_{\rm ss}(X)\approx &-\beta \free (X)-\sum_{n=1}^\infty \frac{(-\beta)^n}{n!}\langle W^n\rangle^c_{\Delta t}(X)+\mathcal{N}.
\end{align}

\section{Perturbation Analysis}
\label{sec:general}

Non-equilibrium steady states are distinguished from equilibrium states by the existence of non-zero mean currents $J_{\rm ss}$ of some conserved quantities (mass density, momentum, charge, etc.), so that the external field $\field$ conjugate to $J$ does work on a system of volume $V$ at a mean rate $\dot{W} = V\field J_{\rm ss}$. In this section, we show that the cumulant differences $\Delta \langle W^n\rangle^c$ for $n>1$ all vanish in the special case where a set of coarse-grained variables including the relevant $J$'s can be found whose combined dynamics are described by a linear overdamped Langevin equation with additive white noise. This assumption of linearity does not imply a restriction to the linear response regime of near-equilibrium thermodynamics, as we show in the following subsection by computing the deviation of $J_{\rm ss}(E)$ from its linear-response form in terms of the parameters of the linear model. In the final subsection we will examine small perturbations around this regime to see how the cumulant differences grow as nonlinearities are introduced.

For some of the calculations, we will assume that the observable $X$ whose probability is being computed is one of the currents $J$. This assumption can be relaxed by a simple change of variables, whose Jacobian will be absorbed into the equilibrium term $\free (X)$, as long as the dynamics of the current still satisfy the given requirements. 

\subsection{Linear Regime}
To compute the conditional averages $\langle \cdot\rangle_{{\rm ss}\to X}$, we start by writing down an equation of motion for the observables in our system, such that the trajectories of the observables for a given realization of the noise can be found by solving the equation and applying the \emph{final} condition that the trajectory \emph{ends} in $X$ at $t = 0$. We will allow for an arbitrary number of observables, contained in a vector $\mathbf{X}$, but will allow for only one current $J$ with steady-state value $J_{\rm ss}$ that is responsible for the steady-state work. This restriction can be relaxed without affecting the final result, but it simplifies the intermediate notation.

We start with the case of a linear Langevin equation for the fluctuation dynamics:
\begin{align}
\label{multieq}
  \dot{\mathbf{X}} = A(\field)\mathbf{X}+B(\field)\xi(t)
\end{align}
where we have defined $\mathbf{X}$ such that $\mathbf{X}=0$ is the most probable value in the steady state. We will choose the first element of $\mathbf{X}$ to contain the current, so that $X_1 = J-J_{\rm ss}$. The matrices $A$ and $B$ are constant in time, but depend on the strength of the external field $\field$.  $\xi(t)$ is a vector of Gaussian white noise, characterized by its mean $\langle \xi_i(t)\rangle = 0$ and two-point function $\langle \xi_i(t)\xi_j(t')\rangle = \delta(t-t')\delta_{ij}$. 

With the ansatz $\mathbf{X}= e^{At}[\mathbf{X}(0) + \mathbf{f}(t)]$, we find
\begin{align}
  A\mathbf{X} + e^{At}\frac{d\mathbf{f}}{dt}&= A\mathbf{X} + B\xi(t)\nonumber\\
  \frac{d\mathbf{f}}{dt} &= e^{-At}B\xi(t)\nonumber\\
  \mathbf{f}(t) &= -\int_t^0 dt'e^{-At'}B\xi(t')
\end{align}
so that $\mathbf{X} = e^{At}\mathbf{X}(0) -\int_t^0 dt'\, e^{A(t-t')}B\xi(t')$ (keeping in mind that $t<0$, since we are specifying the \emph{final} condition $\mathbf{X}(0)$).

Since the first observable $X_1$ is equal to the deviation $J-J_{\rm ss}$ from the mean steady state current, the work done over a given trajectory $\mathbf{X}(t)$ is
\begin{align}
\label{workdef}
  W &=  V \field \int_{-\Delta t}^0 dt [X_1(t)+J_{\rm ss}]\\
&=V\field \int dt\left[ (e^{At})_{1j} X_j(0)+(e^{At})_{1j}f_j(t)+J_{\rm ss}\right],
\end{align}
where we are implicitly summing over all the $j$'s, using the Einstein summation convention.

We can now compute $\Delta\langle W\rangle(X_i(0))$ as a function of any one of the parameters $X_i(0)$, by averaging $W$ over paths that start in the steady state at time $t = -\Delta t\to -\infty$, and end at specified values of $X_i(0)$ at time $t=0$:
\begin{align}
\label{Wrenpert}
\Delta\langle W\rangle(X_i(0)) &= \langle W\rangle_{{\rm ss} \to X_i(0)} - \langle W\rangle_{{\rm ss} \to 0} \nonumber \\
&=V\field\int_{-\infty}^0 dt (e^{A(\field)t})_{1j} \langle X_j(0)\rangle_{X_i(0)}.
\end{align}

To obtain the higher cumulants, we first use the fact that the solution $\mathbf{X}(t)$ obtained above is a sum of independent Gaussian random variables, to conclude that the steady-state distribution for this linear relaxation is itself a Gaussian, which can be written as
\begin{align}
p_{\rm ss}(\mathbf{X}) \propto \exp\left(-\frac{1}{2}\sum C^{-1}_{jk}X_j X_k\right)
\end{align}
where $C$ is the covariance matrix.

We can similarly show that the work distribution is Gaussian. Since all cumulants beyond the second are zero for a Gaussian distribution, we need only compute the contribution from the variance:
\begin{align}
\Delta \langle W^2\rangle^c=& \langle W^2\rangle^c_{{\rm ss} \to X_i} -  \langle W^2\rangle^c_{{\rm ss} \to 0}  \nonumber\\
=& V^2 \field^2\int_{-\infty}^0dt' \int_{-\infty}^0dt\, (e^{At})_{1j}(e^{At})_{1k} \nonumber\\
&\times\left(\langle X_j(0)X_k(0)\rangle^c_{X_i} -\langle X_j(0)X_k(0)\rangle^c_{0}\right).
\label{Phirengauss}
\end{align}
To compute the covariances in the above expression, we need to take a slice through the Gaussian steady-state distribution at the indicated fixed values of $X_i$. The distribution of the remaining variables $\mathbf{X}'$ in this slice is:
\begin{align}
p_{\rm ss}(&\mathbf{X}'|X_i) \nonumber\\
&\propto \exp\left(-\frac{1}{2}\sum_{j,k\neq i} C^{-1}_{jk} X_j X_k -\sum_{j\neq i}C^{-1}_{ij}X_i X_j\right)\\
&= \exp\left(-\frac{1}{2}\sum_{jk}C'^{-1}_{jk}X'_j X'_k - \sum_j \mu_j X'_j\right)
\end{align}
The mean of the new distribution depends on the vector $\mu_j = C^{-1}_{ij}X_i$ and the new covariance matrix $C' = \langle X_j(0)X_k(0)\rangle^c_{X_i} $ is found by removing the $i$th row and column from $C^{-1}$ and then inverting it. The key property of this distribution for our purposes is that $C'$ is independent of the value of $X_i$. It does depend in general on our choice of which row and column to remove, but it does not change when we vary $X_i$ from 0 to some other value.  Applying this fact to equation (\ref{Phirengauss}), we see that the term in parentheses equals zero for all values of $X_i$, and so equation (\ref{pss}) becomes:
\begin{align}
\label{pssgauss}
\ln p_{\rm ss}(X_i) &=  -\beta [\free (X_i) -\Delta\langle W\rangle(X_i)]+\mathcal{N}.
\end{align}

For this strictly linear case, then, the $\Delta \langle W^n\rangle^c$ vanish for all $n>1$. This remains true even if the dynamics are not Markovian for the current $J$ on its own (as is clearly the case in panel (a) of Figure \ref{fig:method} below), as long as there exists a set of observables $\mathbf{X}$ with sufficiently many additional degrees of freedom that dynamics of the form given in equation (\ref{multieq}) apply.

Equation (\ref{pssgauss}) is equivalent to the McLennan ensemble as presented in equation (3.13) of \cite{Komatsu2009} (our averages are evaluated under the driven dynamics, instead of the undriven dynamics, but that change is $O(\epsilon^2)$ in their notation). But as we will show below, the derivation we have provided can extend the range of validity for this representation beyond the linear-response regime to which it has been traditionally restricted. 

This equation also looks similar to the result of macroscopic fluctuation theory that demonstrates the equality of the log of the steady-state distribution for the densities and the ``excess work'' \cite{Bertini2012}. There is no obvious relationship with this result, however, because we have defined $\Delta \langle W\rangle(X_i)$ by subtracting a constant from the bare mean work, rather than decomposing the current into symmetric and asymmetric parts.

\subsection{Departure from Linear-Response $J_{\rm ss}(E)$}
\label{sec:fluct-resp-relat}

In the one-dimensional case, where $X= J-J_{\rm ss}$, the dependence of $J_{\rm ss}(\field)$ on $\field$ is fully determined by the microscopic reversibility assumption underlying (\ref{pssgauss}) combined with the linear equation of motion (\ref{multieq}) and the definition of work (\ref{workdef}). The result is 
\begin{align}
  \label{Jss}
  J_{\rm ss}(\field) = \frac{\beta V\field}{A(\field)}\langle J^2\rangle_{\rm eq}.
\end{align}

We can compare this to the prediction of the Green-Kubo formula of near-equilibrium linear response theory (cf. \cite{Evans2008}):
\begin{align}
  \label{GK}
  J_{\rm ss}^{\rm LR} (\field)= \beta V\field \int_0^\infty dt \langle J(t)J(0)\rangle_{\rm eq}.
\end{align}

Under the linear overdamped Langevin dynamics analyzed above, this becomes:
\begin{align}
  \label{linresp}
   J_{\rm ss}^{\rm LR}(\field) = \frac{\beta V\field}{A(0)}\langle J^2\rangle_{\rm eq},
\end{align}
we see that the system departs from the linear-response regime when the damping rate $A$ begins to depart from its equilibrium value $A(0)$. The size of the departure is given by
\begin{align}
  \label{eq:lindiff}
  \frac{J_{\rm ss}-J_{\rm ss}^{\rm LR}}{J_{\rm ss}^{\rm LR}} = \frac{A(0)}{A(\field)}-1.
\end{align}

Equations (\ref{pssgauss}), (\ref{multieq}) and (\ref{workdef}) also impose a definite relationship between $A$ and $B$:
\begin{align}
  \label{eq:einstein}
  \frac{B(\field)^2}{2A(\field)} = \langle J^2\rangle_{\rm eq}.
\end{align}
When $\field = 0$, this is an analogue to the Einstein relation, constraining the relationship between the effective ``mobility'' and effective ``diffusion coefficient'' for the fluctuations of the current in equilibrium. Equation (\ref{eq:einstein}) says that this relationship continues to hold beyond equilibrium, even where equation (\ref{eq:lindiff}) indicates a large difference between the actual current and the predictions of linear response theory, as long as the equation of motion remains linear. This relationship guarantees that the variance of the steady state distribution remains equal to its equilibrium value in this regime, even as the external drive alters the rate of relaxation.  A related constraint also exists in the multi-dimensional case, where more parameters come into play, and the whole covariance matrix of the steady-state distribution must remain equal to the equilibrium matrix. 

\subsection{First Correction}
\label{sec:pert}
We now allow nonlinear terms to be introduced into the coarse-grained equations of motion. Our first goal is to obtain an expression in terms of work and temperature for the dimensionless expansion parameter that measures the significance of the nonlinearities in the computation of the steady-state distribution via equation (\ref{pss}). We also seek to determine which terms from the cumulant expansion are present at first order in this parameter. For this calculation, we again focus on the 1-D case, where the dynamics of the current $J$ are Markovian. We begin by adding a small nonlinear term added to the linear dynamics for $X = J- J_{\rm ss}(\field)$:
\begin{align}
\label{1deq}
\dot{X} = A(\field)X + \frac{\param}{2}X^2+ B(\field)\xi(t)
\end{align}
where $\xi(t)$ is again a Gaussian white noise term with mean 0 and autocorrelation function $\langle \xi(0)\xi(t)\rangle = \delta(t)$. Note that the sign on $A$ is positive, because we need to average over trajectories whose final conditions are specified, as opposed to the usual initial conditions. The coefficient $\param$ has dimensions of 1/[time][current], and part of our task is to convert it into a dimensionless expansion parameter.%

We can write the solution as a power series in $\param$:
\begin{align}
X(t) = X^{(0)}(t) + \param X^{(1)}(t)+\param^2 X^{(2)}(t)+\dots.
\end{align}
Plugging this in to equation of motion and collecting terms in powers of $\param$, we find
\begin{align}
\dot{X}^{(0)} &= aX^{(0)}+B\xi(t)\\
\dot{X}^{(1)} &= aX^{(1)}+\frac{1}{2}(X^{(0)})^2.
\end{align}
The first is the same as the equation we solved above, giving
\begin{align}
X^{(0)}(t) = e^{At}(X(0)+f(t))
\end{align}
with $f(t) =- \int_t^0 e^{-At'}B\xi(t')dt'$.  We can use this result in the second equation and solve in a similar way to obtain
\begin{align}
X^{(1)}(t) = -\int_t^0 dt' \,e^{A(t-t')}\frac{(X^{(0)}(t'))^2}{2}.
\end{align}
Integrating the perturbative solution $X(t) = X^{(0)}(t)+\param X^{(1)}(t)+ O(\param^2)$, we obtain the work done for a given realization of the noise:
\begin{align}
W& =V \field\int_{-\Delta t}^0dt \Big[ e^{At}X(0)+e^{At}f(t) \nonumber\\
&\qquad\qquad\qquad {}+\param\int_t^0 dt'\,e^{A(t-t')}\frac{(e^{At'}X(0)+e^{At'}f(t'))^2}{2}\nonumber\\
&\qquad\qquad\qquad {}+O(\param^2) +J_{\rm ss}\Big] \\
&= \frac{ V\field}{A} X(0) + \frac{V\field\param }{4 A^2}X(0)^2+V \field\int_{-\infty}^0 dt\Big\{ e^{At}f(t) \nonumber\\
&+\int_t^0 dt' \,e^{A(t+t')}[ \param X(0)f(t')+\frac{\param}{2}f(t')^2]+O(\param^2) +J_{\rm ss}\Big\}
\label{eq:work}
\end{align}

Now we need to use this expression to compute the $\Delta\langle W^n\rangle^c(X)$. Since $f(t)$ is independent of the choice of ending state $X(0)$, we find
\begin{align}
\label{wmain}
\Delta\langle W\rangle(X) &= \frac{ V\field}{A(\field)} X + \param\frac{V\field}{4A(\field)^2}X^2+O(\param^2)\nonumber\\
&= \Delta \langle W\rangle^{(0)}(X)+\param\frac{V\field}{4A(\field)^2}X^2+O(\param^2),
\end{align}
where $\Delta \langle W\rangle^{(0)}(X)$ is computed under the linear dynamics alone. The quadratic term in equation (\ref{wmain}) provides a correction to the equilibrium variance at order $\param$, which also breaks the relationship (\ref{eq:einstein}) between $A(\field)$ and $B(\field)$.

To compute the higher cumulants, we first note that the first two terms in equation (\ref{eq:work}) are not random variables, and serve as a constant offset that makes no contribution to the cumulants of order 2 and higher. Focusing on the term in curly brackets, then, we see that the $\param f(t')^2$ term can only be part of an $X(0)$-dependent term in $\Delta \langle W^n\rangle^c$ when it is multiplied by some nonzero power of the $\param X(0)f(t')$ term. It therefore contributes only to the $O(\param^2)$ part of the expression and to the overall normalization. The remaining part of the work can be expressed as a sum of independent Gaussian random variables $\xi(t)$, so it is itself a Gaussian random variable. This implies that it has no nonzero cumulants beyond the variance, so that all higher cumulants are $O(\param^2)$. The variance is given by
\begin{align}
\label{varnow}
 \langle W^2\rangle^c_{{\rm ss}\to X}&= \langle W^2\rangle_{{\rm ss}\to X}-\langle W\rangle^2_{{\rm ss}\to X}\nonumber\\
& =  2V^2\field^2\param\int_{-\infty}^0dt\int_{-\infty}^0 dt'\int_{t'}^0 dt''\, e^{A(t+t'+t'')}X\nonumber\\
&\,\,\,\,\,\,\,\,\times\langle f(t)f(t'')\rangle +O(\param^2)+\mathcal{N}
\end{align}
where we have absorbed the $X$-independent terms into $\mathcal{N}$.

To simplify this, we must compute the autocorrelation function of $f(t)$. We first consider the case $|t''|\geq|t|$:
\begin{align}
  \langle f(t)f(t'')\rangle &= B^2 \int_t^0 ds\int_{t''}^0 du \,e^{-A(s+u)}\langle \xi(s)\xi(u)\rangle\nonumber\\
&= B^2 \int_t^0 ds \int_{t''}^t du \,e^{-A(s+u)} \delta(s-u)\nonumber\\
&\,\,\,\,\,\,\,+ B^2\int_t^0 ds \int_t^0 du\, e^{-A(s+u)} \delta(s-u)\nonumber\\
&= B^2 \int_t^0 ds \,e^{-2As} = \frac{B^2}{2A}(e^{-2At}-1).
\end{align}
If $|t''|\leq |t|$, this becomes 
\begin{align}
  \langle f(t)f(t'')\rangle = \frac{B^2}{2A}(e^{-2At''}-1).
\end{align}
Combining the two answers, we find
\begin{align}
\langle f(t)f(t'')\rangle = \frac{B^2}{2A}(e^{-2At_m}-1)
\end{align}
where $t_m$ is equal to whichever of $t,t''$ has the smaller absolute value.

Plugging this in to equation (\ref{varnow}), we find
\begin{align}
 \langle W^2\rangle^c_{X}=& 2V^2\field^2\param X\frac{B^2}{2A}\nonumber\\
&\times\int_{-\infty}^0dt\int_{-\infty}^0 dt'\int_{t'}^0  dt''\, (e^{A(t'-|t-t''|)} \nonumber\\
&\,\,\,\,\,\,\,\,\,\,\,\,\,\,\,\,\,\,\,\,- e^{A(t+t'+t'')}) +O(\param^2)+\mathcal{N}\nonumber\\
=& \param \frac{V^2\field^2XB^2}{A^4}+O(\param^2)+\mathcal{N}.
\end{align}

Thus we obtain the next term in the cumulant expansion in equation (\ref{pss}):
\begin{align}
\frac{\beta^2}{2}\Delta \langle W^2\rangle^c(X)&= \frac{\beta^2}{2}(\langle W^2\rangle_{{\rm ss}\to X}^c - \langle W^2\rangle_{{\rm ss}\to 0}^c)\nonumber\\
&= \param\frac{\beta^2 V^2 \field^2B^2}{2A^4}X+O(\param^2).
\end{align}
By the argument given above equation (\ref{varnow}), the rest of the terms in the cumulant expansion are guaranteed to be $O(\mu^2)$. To obtain the correct dimensionless form of $\param$, we need to compare this term to the $\beta \Delta \langle W\rangle$ term, and see what combination of parameters controls their relative size. Using equation (\ref{wmain}), we find
\begin{align}
\frac{\beta^2}{2}\Delta \langle W^2\rangle^c(X)&= \param\frac{\beta V\field B^2}{2A^3}\beta \Delta\langle W\rangle(X)+O(\param^2),
\label{phimain}
\end{align}
so that $\tilde{\param} = \param\frac{\beta V\field B^2}{2A^3}$ is the appropriate dimensionless version of $\param$ that controls how quickly the expansion converges. 

We can interpret $\tilde{\param}$ thermodynamically by comparing to the quadratic $O(\param)$ term in (\ref{wmain}) that corrects the variance of the distribution. Using the fact that the variance in the unperturbed steady-state distribution is $\sigma_X^2= B^2/2A$ (cf. eq. 3.8.74 and 4.3.23 in \cite{Gardiner}), we can write
\begin{align}
\label{thermpar}
\tilde{\param}= 4\beta[\Delta\langle W\rangle(\sigma_X)-\Delta \langle W\rangle^{(0)}(\sigma_X)]
\end{align}
which is four times the typical extra mean work difference due to the nonlinear term in the dynamics, divided by $k_BT$.  

Thus we conclude that the first terms in the expansion of $p_{\rm ss}(X)$ about the linearized dynamics in this 1-D case are given by
\begin{align}
\ln p_{\rm ss}(X) &=  -\beta [\free (X) -\Delta\langle W\rangle(X)]-\frac{\beta^2}{2} \Delta \langle W^2\rangle^c(X) \nonumber\\
\label{psspert}
&+ O(\tilde{\param}^2)+\mathcal{N}
\end{align}
with $\tilde{\param}$ as defined above. 

In the multidimensional case, defining the $\tilde{\param}$ explicitly in terms of the model parameters is more challenging, but it is reasonable to suppose that the thermodynamic expression (\ref{thermpar}) should still give a good estimate of its size. In the next section, we will apply this result to our numerical simulation of a sheared colloid, which will require the finite-time approximate form
\begin{align}
\ln p_{\rm ss}(X) & \approx -\beta [\free (X) -\langle W\rangle_{\Delta t}(X)]-\frac{\beta^2}{2} \langle W^2\rangle^c_{\Delta t}(X) \nonumber\\
\label{psspert2}
&+ O(\tilde{\param}^2)+\mathcal{N}
\end{align}
which holds when $\Delta t\gg \tau = 1/A$, with the finite-time cumulants defined in equation (\ref{sswork}).

\section{Shear Thinning Example}
\label{sec:system}
Strongly driven colloids are commonly used as examples of far-from-equilibrium steady state systems where the relationship between currents and applied fields can violate the predictions of linear response theory (cf. \cite{Fuchs2005,Harada2009,Kruger2009,Speck2009,Zhang2011}). Using the shear stress $\sigma_{xy}$ as the current and minus the shear rate $-\dot{\gamma}$ as the conjugate field, one can describe departures from linear response in terms of the dependence of the viscosity $\eta = -\sigma_{xy}/\dot{\gamma}$ on $\dot{\gamma}$. In the linear response regime, $\sigma_{xy}$ is proportional to $\dot{\gamma}$, and $\eta$ is constant. As the shear rate is increased in a typical colloid, $\sigma_{xy}(\dot{\gamma})$ becomes sublinear, so $\eta$ decreases and the suspension is said to ``shear thin'' (cf. \cite{Brader2010}). In this section, we describe how to measure $\langle W\rangle_{\Delta t}(\sigma_{xy})$ and $\frac{\beta^2}{2} \langle W^2\rangle^c_{\Delta t}(\sigma_{xy})$ in computer model of a colloid that exhibits shear thinning. We will show that the above expansion in the degree of nonlinearity converges quickly in this case, so that the $\tilde{\param} \to 0$ form given in equation (\ref{pssgauss}) generates a qualitatively correct description of the shear thinning phenomenon, while the $O(\tilde{\param})$ correction in equation (\ref{psspert2}) is sufficient to maintain quantitative agreement with the actual distribution well into the thinning regime.

\subsection{Setup}
Consider a suspension of small identical spheres in a Newtonian solvent at a fixed temperature. The particles are small enough that Brownian motion can equilibrate their spatial configuration rapidly compared with the duration of a typical experiment, producing a steady state independent of initial conditions. Electrostatic repulsion keeps the spheres far enough apart that the disturbance each particle creates in the flow field has no effect on the trajectories of the other particles, while ions in the solvent screen the charges and exponentially suppress the interaction at large separations. 

\begin{figure}
  \includegraphics[width=7cm]{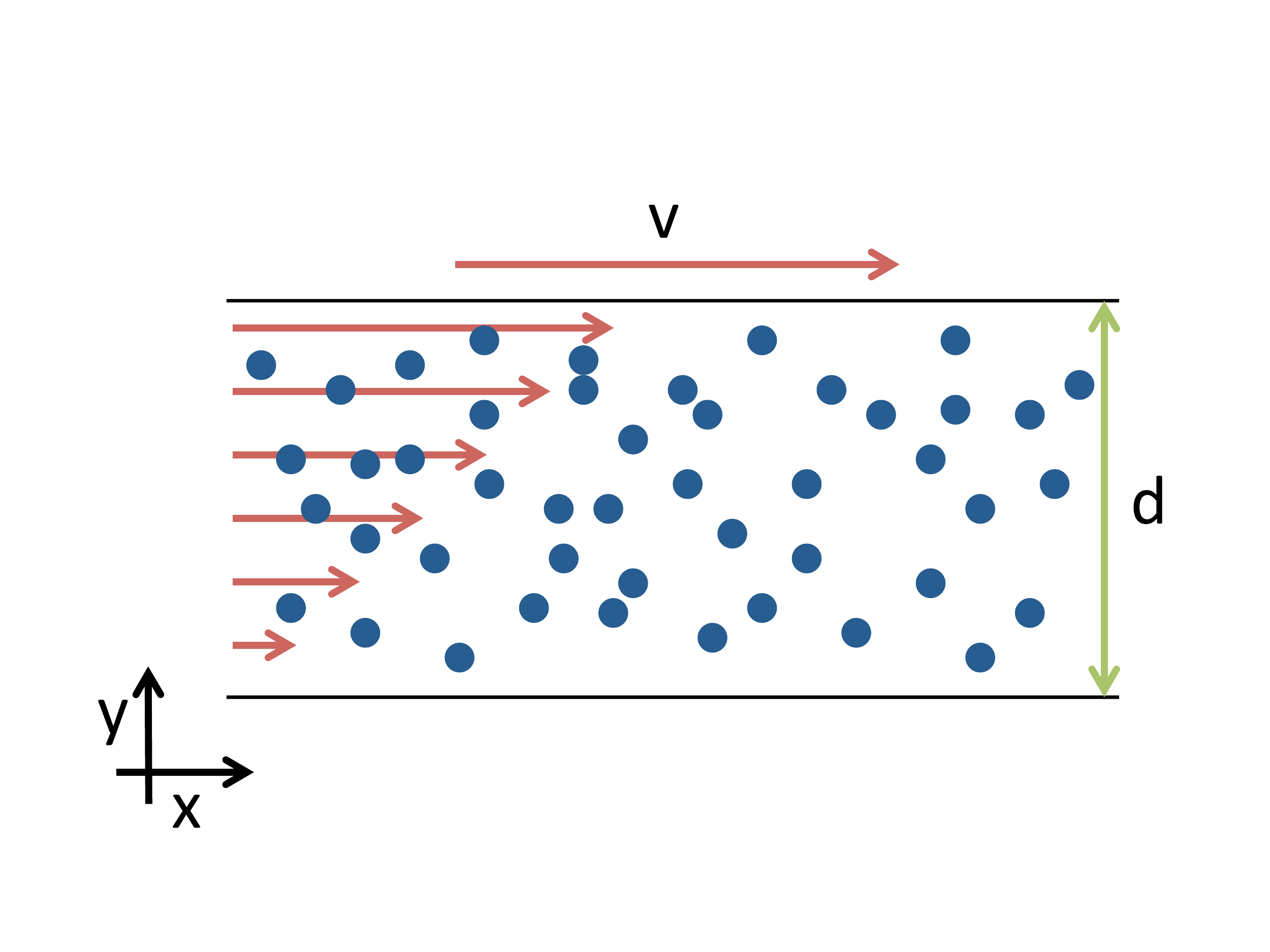}
  \caption{(Color online) Shear cell with periodic boundary conditions along flow direction. The reflecting walls on the top and bottom are separated by a distance $d$. The top wall moves at constant speed $v$ in the $x$ direction, while the bottom wall is fixed, causing a linear gradient $\dot{\gamma} = v/d$ in the solvent flow velocity along the $y$ direction.}
  \label{fig:setup}
\end{figure}

A nonequilibrium steady state can be created by moving one wall of the chamber containing the suspension at a constant velocity $v$  while keeping the opposite wall fixed, thus setting up a steady shear flow in the gap of width $d$ between the walls. (The wall velocity is ultimately due to some time-varying fields $\control(t)$ - like the fields inside an electric motor). The strength of the shear flow can be quantified in a form independent of the system dimensions as the ``shear rate'' $\dot{\gamma} = v/d$. A constant shear rate can be maintained by using periodic boundary conditions in the flow direction (which can be approximated in experiment by using a cylindrical geometry). As indicated in Figure \ref{fig:setup}, we will define coordinates such that the moving wall travels in the $+x$ direction and the $y$ axis points from the stationary wall to the moving wall.

Two important dimensionless parameters for the dynamics of a sheared colloid are the Reynolds number Re = $\rho\dot{\gamma}a^2/\eta_0$ and the Peclet number Pe = $\dot{\gamma}a^2b/k_B T$.
Here $\rho$ is the mass density of the fluid (assumed to be comparable to the density of the particles), $a$ is the radius of a particle, $\eta_0$ is the viscosity of the suspending fluid, $b$ is the drag coefficient of a particle $(=6\pi\eta_0 a$ for a sphere with no-slip boundary conditions), $k_B$ is Boltzmann's constant, and $T$ is the temperature of the heat bath coupled to the fluid.
Re measures the importance of inertia relative to viscous drag, and Pe measures the importance of motion by convection in the shear flow relative to diffusive motion in a dilute suspension (when the suspension becomes sufficiently dense, the equilibrium relaxation is slowed down by the interactions among the particles, and the relevant dimensionless parameter becomes the density-dependent Weissenberg number). Pe thus measures the distance from equilibrium, so that Pe $\ll 1$ gives rise to the linear-response regime, and Pe $\gg 1$ constitutes the ``far from equilibrium'' regime where shear thinning occurs. 

In the overdamped limit where Re $\ll 1$, the instantaneous velocity of the particles can be regarded as fully determined by their spatial configuration (up to the rapidly equilibrating contribution from Brownian motion), so the set of particle positions is sufficient to define the full microstate, and the corresponding simulation algorithm becomes very simple. Re can be kept in this regime while sweeping Pe up to any desired maximum value Pe$_{\rm max}$ by choosing a viscosity such that $\eta_0\gg \sqrt{\rho {\rm Pe}_{\rm max}k_B T/a}$. 


\subsection{Equations of Motion}
To describe this system mathematically, we will use the model employed in \cite{Speck2009,Zhang2011} for the investigation of departures from near-equilibrium linear-response behavior in nonequilibrium steady states. This model can be numerically simulated with the dilute limit of the Brownian Dynamics of Ermak and McCammon \cite{Ermak1978} or of the Stokesian Dynamics of Brady and Bossis \cite{Brady1988}, where hydrodynamic interactions are ignored, and the Re $\ll 1$ limit is invoked so that the particle inertia becomes irrelevant.  These assumptions lead to the following discretized equations of motion for the position $(x_i,y_i)$ of particle $i$ confined to move in two dimensions:
\begin{align}
\label{x}
  x_i(t+\Delta t) &= x_i(t) + y_i(t)\dot{\gamma}\Delta t + \frac{1}{b}\sum_{j} \hat{\mathbf{x}}\cdot \mathbf{F}_{ji}\Delta t+\Delta x^r_i\\
\label{y}
  y_i(t+\Delta t) &= y_i(t) + \frac{1}{b}\sum_{j} \hat{\mathbf{y}}\cdot \mathbf{F}_{ji}\Delta t + \Delta y^r_i
\end{align}
where $b$ is the drag coefficient for the particles, and $\mathbf{F}_{ji}$ is the force exerted on particle $i$ by particle $j$. We choose the force to be a screened Coulomb repulsion, with potential energy $U(r) = k_BT e^{-r/\lambda}zl_B/r$ as a function of the distance $r$ separating a pair of particles. $\lambda$ is the screening length, $l_B$ is the Bjerrum length, and $z$ is the number of elementary charges on each particle.  $\Delta x^r_i$ and $\Delta y^r_i$ are random displacements due to Brownian motion. Brownian and Stokesian dynamics assume that the solvent degrees of freedom always equilibrate before the spatial configuration of the colloidal particles can change appreciably, so that the Brownian displacements can be chosen from a Gaussian distribution whose variance $(2k_BT/b)\Delta t$ is related to the temperature and drag coefficient by the Einstein relation. Equations (\ref{x}) and (\ref{y}) are then simply iterated with a small enough time step that the results are insensitive to variations in time-step size \cite{supp2}. 

As mentioned at the beginning of this section, we are considering the case where the particle size is much smaller than $\lambda$ or $zl_B$, so that ``hydrodynamic interactions'' (particle-particle interactions mediated by disturbances in the solvent flow) make a negligible impact on the particle trajectories. This is what allows us to use the ``dilute limit'' of the Stokesian or Brownian Dynamics, where the mobility and resistance tensors are diagonal and independent of particle positions. Another consequence of this limit is that the actual particle radius $a$ does not appear in the equations of motion; we therefore use the screening length $\lambda$ as the microscopic length scale for computing the Peclet number and measuring distance from equilibrium.

\subsection{Shear Stress}

The macroscopic viscosity of the whole suspension at equilibrium will be larger than $\eta_0$, because both the disturbance of the flow field produced by individual particles and the mutual repulsion between pairs of particles make the suspension harder to shear than the bare fluid. As the suspension is sheared, however, the contribution of the particle repulsion to the viscosity decreases, and the suspension shear thins (cf. \cite{Brader2010}). The particles cause the shear stress to vary with position in the suspension, so we define an overall shear stress for the system by averaging the local shear stress at the moving wall of the system over the whole wall area. This will be convenient for computing the work done by the moving wall later on, and gives us a macroscopic parameter that can be directly observed in experiment via a measurement of the force applied to the wall. As shown in the Appendix, for a suspension of particles in a Newtonian solvent in the limit of zero Re with no hydrodynamic interactions, the instantaneous mean shear stress $\sigma^{\rm wall}_{xy}$ exerted by the fluid on the moving wall is:
\begin{align}
\label{sigmafull}
\sigma_{xy}^{\rm wall} = \sigma_{xy}^I + \sigma_{xy}^0.
\end{align}
The first term depends on the force $\mathbf{F}_{ij}$ exerted by each particle $i$ on the other particles $j$:
\begin{align}
\label{sigmai}
\sigma_{xy}^I = \frac{1}{2V} \sum_{i\neq j} \mathbf{\hat{x}}\cdot \mathbf{F}_{ij} \Delta y_{ij}.
\end{align}
Here $V$ is the system volume, $\mathbf{\hat{x}}$ is the unit vector in the $+x$ direction, and $\Delta y_{ij} = y_j-y_i$. The right-hand side can be unambiguously determined from the system microstate, which we are taking to be the list of positions of all the particles. 
We can therefore choose $X = \sigma_{xy}^I$ as our coarse-grained observable and test how well the first terms in our expansion describe its fluctuations at a fixed shear rate $\dot{\gamma}$.  The remaining term $\sigma_{xy}^0$ is independent of the particle positions, so the work done by the moving wall will depend on the particle configuration through $\sigma_{xy}^I$ alone. 

When the shear rate is small compared to the diffusive relaxation rate, the overall viscosity of the suspension can be computed from the equilibrium fluctuations in $\sigma_{xy}$ using linear response theory \cite{Evans2008}. As the shear rate continues to increase, the viscosity begins to deviate from this value as the suspension shear thins. In the following sections, we will use equation (\ref{psspert2}) to determine the most probable value of $\sigma_{xy}^I$ and hence the contribution $\sigma_{xy}^I/\dot{\gamma}$ to the viscosity in both the linear response regime and the shear thinning regime. We start in section \ref{sec:work} by demonstrating how to extract $\langle W\rangle_{\Delta t}(\sigma_{xy}^I)$ and $\frac{\beta^2}{2} \langle W^2\rangle^c_{\Delta t}(\sigma_{xy}^I) $ from the simulation in a way that should also be experimentally accessible. Then in section \ref{sec:results} we plot the results over a range of shear rates, and compare the prediction of equation (\ref{psspert2}) with the observed steady-state distribution.

\subsection{Measuring the Work Cumulants}
\label{sec:work}
The rate at which the moving wall does work on the fluid
is just the force $-A\sigma^{\rm wall}_{xy}$ it exerts against the fluid (where $A$ is the surface area of the wall) times the speed of the wall $\dot{\gamma}d$. Using equation (\ref{sigmafull}), we thus obtain:
\begin{align}
\label{diss}
  \dot{W}=-V\dot{\gamma}\sigma^I_{xy} + \dot{W}_0.
\end{align}
where $\dot{W}_0$ is the part of the work that does not depend on the configuration of the particles. Since $\dot{W}_0$ only affects the normalization, but not the shape of the distribution, we will set it to zero for the purpose of the calculations in this section. Then we can treat $\sigma_{xy}^I$ as the current $J$ and minus the shear rate $-\dot{\gamma}$ as its conjugate field $\field$.

\begin{figure}
  \centering
  \includegraphics[clip=true, trim = .7cm 0 0 0]{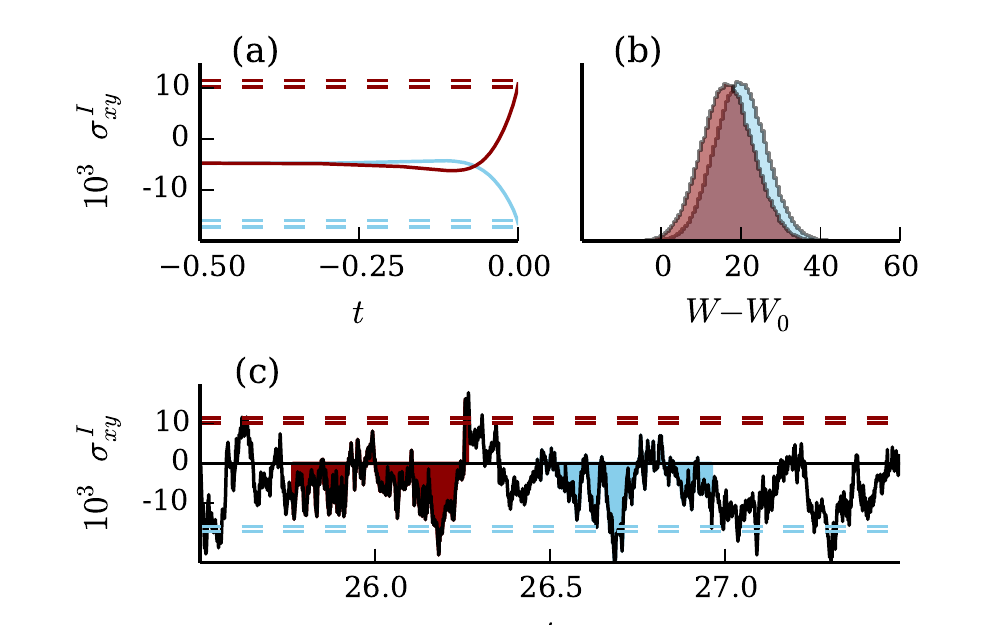}
  \caption{(Color online) (a) $\sigma^I_{xy}(t)$ averaged over all trajectory segments of length $\Delta t = 0.5$ that end at a specified value, from a simulation run with the same parameters as Figure \ref{fig:worklines}, and Pe = 19. Two arbitrarily chosen ending state restrictions are indicated by dotted lines, with the corresponding average trajectories shown in the same colors. (b) Measured probability distributions of work done over all the trajectories that go into the averages of panel (a), shaded in the same colors. (c) A portion of the raw $\sigma^I_{xy}$ timeseries from which the other two panels were generated. The ending state restrictions of panel (a) are indicated with dotted lines of the same color, and two typical trajectory segments ending at these values are shaded in these colors. The shaded areas are proportional to the interaction-dependent part of the work, according to equation (\ref{diss}).}
  \label{fig:method}
\end{figure}

Using equation (\ref{diss}), we can now compute the work done along any stochastic trajectory $\sigma^I_{xy}(t)$ by simply integrating the trajectory with respect to time.  We can then compute the distribution of $W$ for trajectories ending at a given $\sigma_{xy}^I$ value by letting the system relax to the steady state at some value of $\dot{\gamma}$ and run there for a long time, while continuously recording the fluctuations in $\sigma^I_{xy}$ (which can in principle be determined directly from the fluctuations in the force applied to the moving plate in an experiment). As shown in Figure (\ref{fig:method}), we then choose some time interval $\Delta t$ longer than the timescale $\tau$ of relaxation to the steady state, and compute both the work and the final value of $\sigma^I_{xy}$ for every segment of length $\Delta t$ in the whole trajectory. Finally, we bin the work outputs by the corresponding final value of $\sigma^I_{xy}$ to obtain the distribution of work for each bin, from which we can compute the cumulants $\langle W^n\rangle^c_{\Delta t}(\sigma_{xy}^I)$. 

Figure \ref{fig:worklines0} shows the zeroth order term $\langle W\rangle_{\Delta t}(\sigma^I_{xy})$ in the expansion (\ref{psspert2}) and the first-order correction $(\beta/2) \langle W^2\rangle^c_{\Delta t}(\sigma^I_{xy})$ as a function of ending value of $\sigma^I_{xy}$, at three different values of the dimensionless shear rate Pe.

\section{Simulation Results}
\label{sec:results}
With this procedure in hand for extracting $\langle W\rangle_{\Delta t}$ and $\langle W^2\rangle^c_{\Delta t}(\sigma^I_{xy})$, we can use equation (\ref{psspert2}) to compute the steady-state distribution of $\sigma_{xy}^I$ from the work statistics, and compare it to the directly measured distribution in our simulation. 

We simulated a sheared colloidal monolayer of $N = 100$ particles using the equations of motion (\ref{x}) and (\ref{y}) \cite{supp2}. The colloid was confined to a square box of side length 20, with reflecting boundary conditions on the moving wall and the opposite wall, and periodic boundary conditions on the other sides. This concentration is sufficiently dilute that the equilibrium relaxation time does not vary much with changes in concentration, so the Peclet number should still be the relevant parameter for measuring distance from equilibrium. The other parameters were chosen as $k_BT = b = \lambda = zl_B = 1$. We ran this simulation for 20 different values of Pe, from 0 to 19, generating trajectories with lengths up to $t = 120,000$  in the given units, with time step size 0.001. The simulations were initialized with uniform random distributions of particle positions, and the initial transients were removed from the timeseries before analysis.

To apply equation (\ref{psspert2}), we first need to find the equilibrium free energy $\free (\sigma^I_{xy})$ as a function of $\sigma^I_{xy}$. This can be extracted from the simulation by extracting the distribution of equilibrium fluctuations from a run at $\dot{\gamma} = 0$, taking the natural logarithm, and multiplying by $-k_B T$. 

\begin{figure}
  \centering
  \includegraphics[clip=true, trim = 2.2cm 2.5cm 1.5cm 2cm]{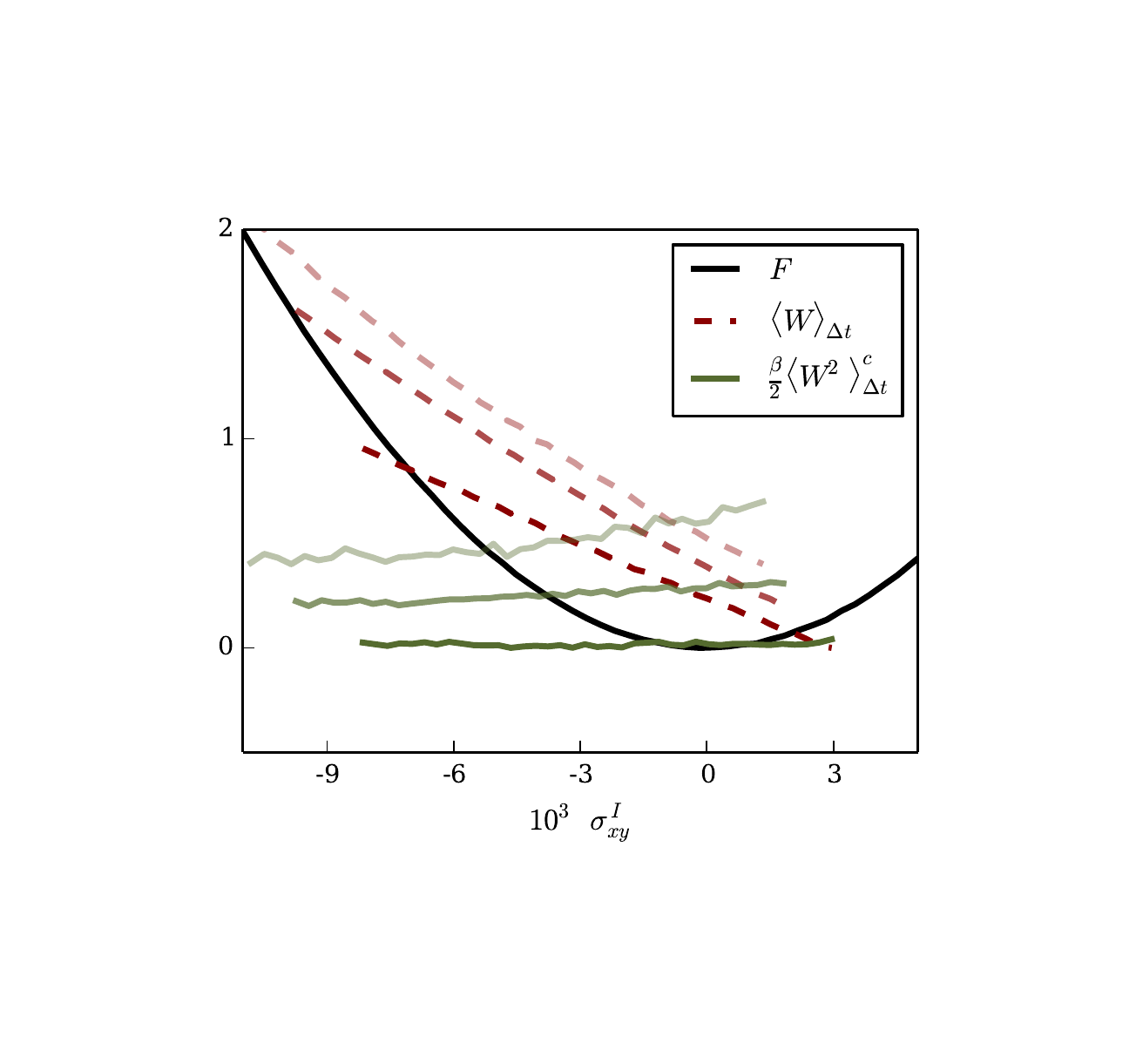}
  \caption{(Color online) Using the method illustrated in Figure \ref{fig:method}, we compute $\langle W\rangle_{\Delta t}(\sigma^I_{xy})$ and $(\beta/2) \langle W^2\rangle^c_{\Delta t}(\sigma^I_{xy})$ for a range of values of $\sigma^I_{xy}$, and plot this data for Pe values 4, 10, and 19, increasing from bottom to top. Also plotted is the free energy $F$ extracted from the Pe = 0 simulation run.}
\label{fig:worklines0}
\end{figure}

Figure \ref{fig:worklines0} shows how $F$ and the other two terms in equation (\ref{psspert2}) depend on $\sigma_{xy}$, with the nonequilibrium terms evaluated at three different values of the shear rate. $F$ is parabolic near $\sigma_{xy}^I = 0$, but requires a fourth-order polynomial to fit the far tails. $\langle W\rangle_{\Delta t}$ starts out linear at low shear rates, starts curving slightly by Pe = 10, and becomes noticeably quadratic by Pe = 19, indicating that the $O(\tilde{\param})$ term has become important. $\langle W^2\rangle^c_{\Delta t}$ is independent of $\sigma_{xy}^I$ at low shear rates, but starts becoming $\sigma_{xy}^I$-dependent at about the same shear rate as $\langle W\rangle_{\Delta t}$ begins to deviate from linearity, as expected from our analysis in section \ref{sec:pert}.

\begin{figure}
  \centering
  \includegraphics[clip=true, trim = 2.4cm 2.5cm 1.5cm 2.5cm]{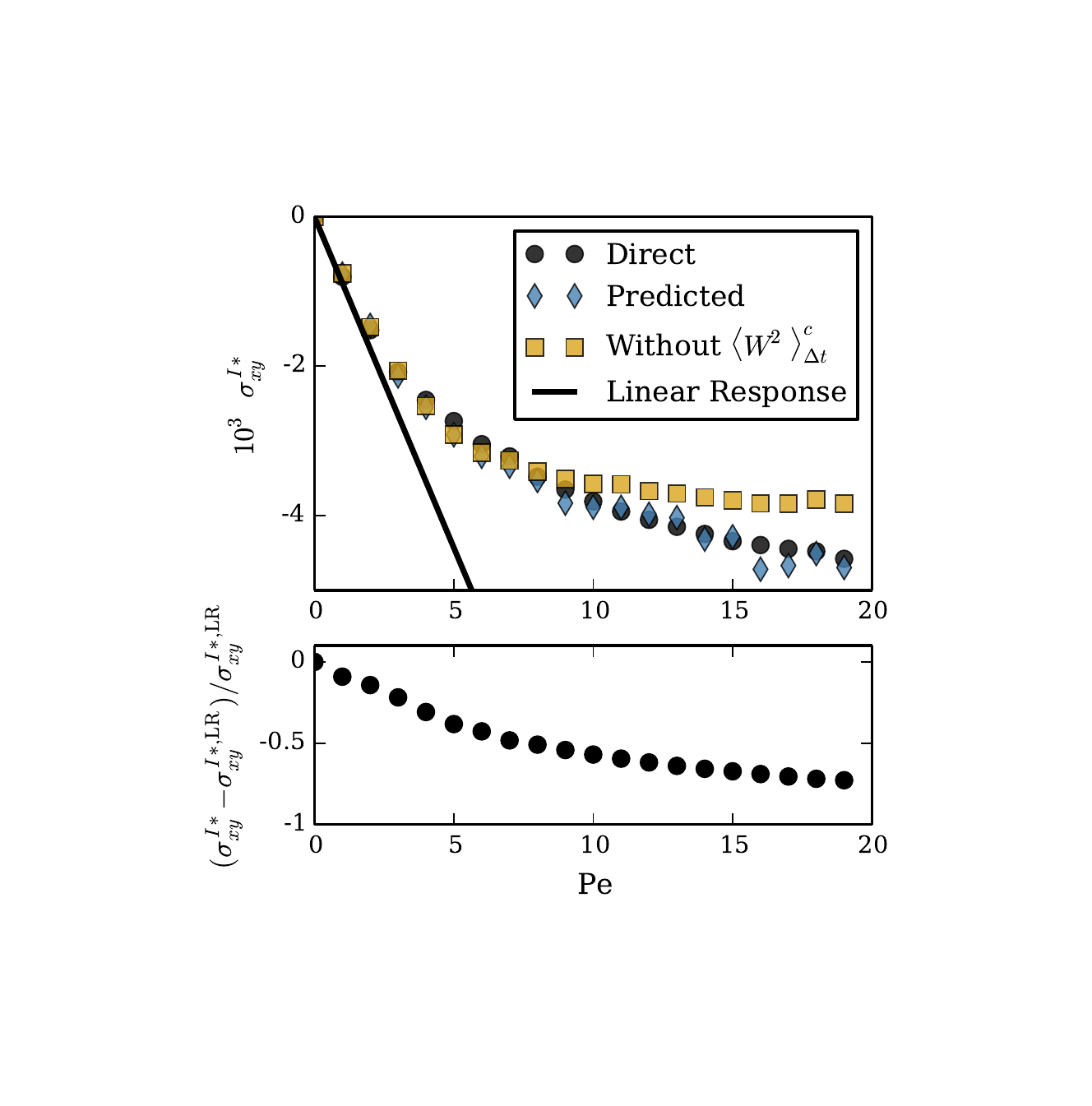}
  \caption{(Color online) Top: Location $\sigma^{I*}_{xy}$ of peak of probability distribution, computed via equation (\ref{psspert2}), compared with directly measured most frequent values in the simulation \cite{supp2}. Also plotted is the prediction using the mean work alone via equation (\ref{pssgauss}), without the $O(\tilde{\mu})$ variance term, as well as the linear response prediction $\sigma_{xy}^{I*,{\rm LR}}$. Bottom: Relative size of departure from linear-response prediction.}
\label{fig:worklines}
\end{figure}

Figure \ref{fig:worklines} shows the location of the peak of the steady-state distribution as a function of Pe computed using equation (\ref{psspert2}), compared with the distribution directly sampled from the simulation. We have also plotted the result ignoring the $\Delta \langle W^2\rangle^c$ term, to show the size of the impact of the $O(\tilde{\param})$ correction compared to the zeroth order expression (\ref{pssgauss}). When we compare both curves to the most probable values of $\sigma^I_{xy}$ actually measured in the simulations, we see that the zeroth-order expression correctly captures the qualitative shear thinning behavior: $\sigma_{xy}^I$ departs from its linear-response dependence on $\dot{\gamma}$ and eventually saturates, causing the interparticle force contribution $ -\sigma_{xy}^I/\dot{\gamma}$ to the viscosity to fall off as $1/\dot{\gamma}$. This approximation predicts that the saturation occurs sooner than it actually does, but the first-order term appears to entirely compensate for the discrepancy. The straight line is the linear-response prediction for the mean shear stress, computed from the equilibrium fluctuations using the Green-Kubo formula given in equation (\ref{GK}). The bottom panel shows the relative difference between the linear-response prediction and the location of the observed peak of the probability distribution, as defined in equation (\ref{eq:lindiff}) of section \ref{sec:general}. 

\begin{figure}
  \centering
  \includegraphics[clip = true, trim = 0 0.7cm 0 1.6cm]{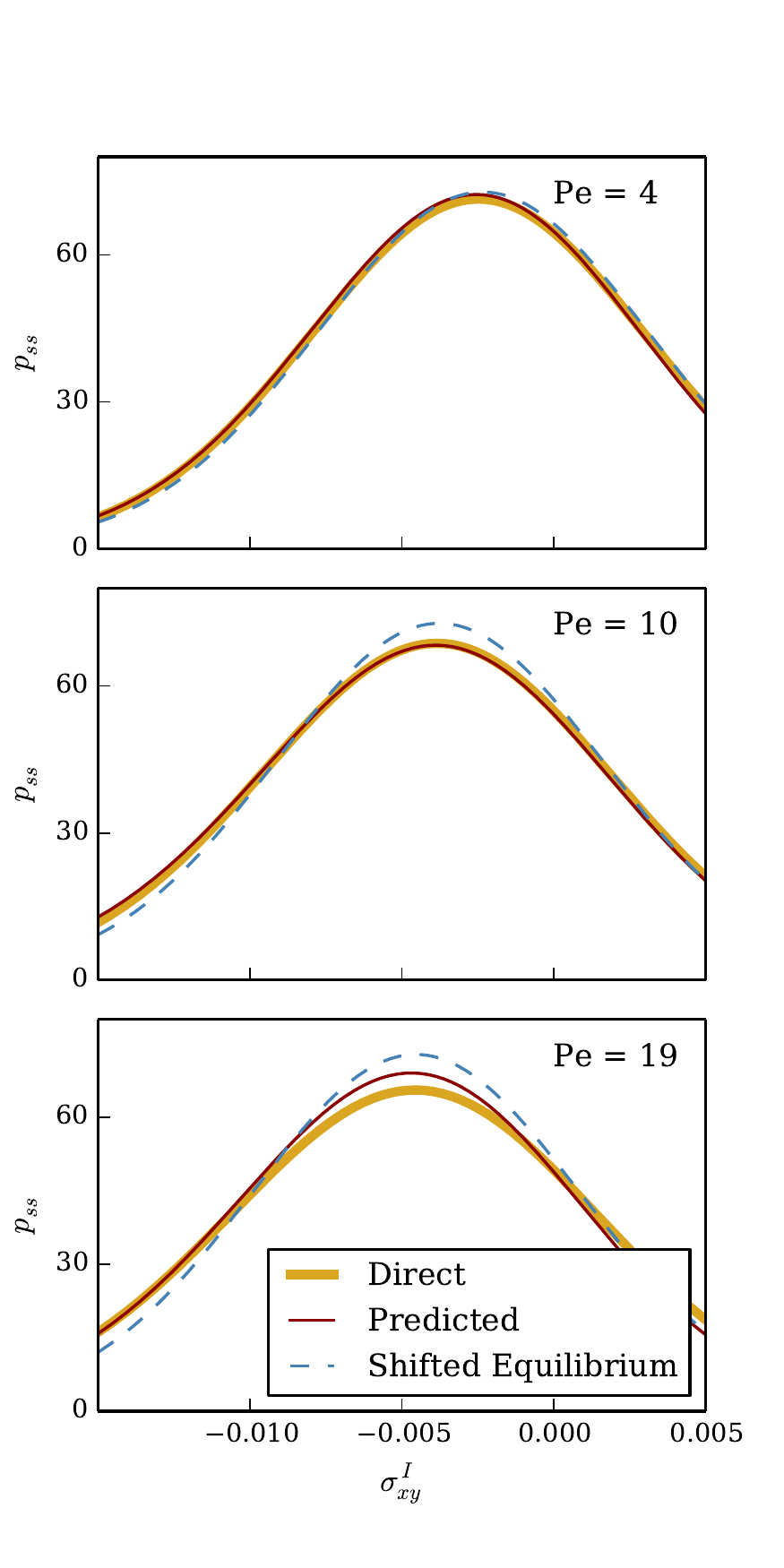}
  \caption{(Color online) Full probability distribution from equation (\ref{psspert2}), compared with shifted equilibrium distribution and with distribution directly sampled from simulation at Pe = 4, Pe = 10 and Pe = 19. }
  \label{fig:dist}
\end{figure}

In Figure \ref{fig:dist} we have plotted the normalized probability distributions at Pe = 4, 10, and 19, showing the direct measurement and the prediction of equation (\ref{psspert2}), smoothed with polynomial fits to the data in Figure \ref{fig:worklines0}. Also plotted is the equilibrium distribution shifted to the location of the new peak, to better visualize the change in the shape of the distribution at high shear rates. The variance is still within 2\% of the equilibrium value at Pe = 4, even though the relative deviation of the mean from the linear-response prediction has already reached 30\%. This is an example of a case where the constraint of equation (\ref{eq:einstein}) derived from the linear approximation remains in effect beyond the linear-response regime.  Equation (\ref{psspert2}) successfully accounts for the change in variance visible at Pe = 10, although it fails to fully capture the asymmetry that enters the distribution at Pe = 19, which should depend on higher-order terms according to the analysis of section (\ref{sec:pert}). 

\section{Discussion}
\label{sec:conclusion}

We have shown that the probability distribution of a macroscopic observable in a driven steady state can be written as a perturbation expansion in the nonlinearities of the coarse-grained dynamics, which can converge quickly in some systems even under strong driving. The approximate formulas (\ref{pssgauss}) and (\ref{psspert}) give a good description of the statistics of the steady state under the following conditions:
\begin{itemize}
\item The transition rates between system states must satisfy the ``microscopic reversibility'' condition given in equation (\ref{crooks1}).
\item The system exponentially loses memory of its initial conditions, with a decay time that is short compared to the duration of the experiment or simulation. 
\item The contribution to the mean work difference $\Delta \langle W\rangle$ (defined in equation (\ref{renormalized})) due to nonlinear terms in the coarse-grained fluctuation dynamics is small compared to the thermal energy scale $k_B T$.
\end{itemize}

The simulation results presented in Section \ref{sec:work} and Section \ref{sec:results} illustrate the physical meaning of the terms in the expansion, and confirm that it can converge rapidly even when external drives are strong enough to make the peak of the steady-state distribution depart significantly from the linear-response prediction.

Evaluating the distribution by measuring $\langle W\rangle_{\Delta t}$ and $\langle W^2\rangle^c_{\Delta t}$ numerically, as done in section \ref{sec:results} is much more computationally expensive than a direct measurement of the distribution from the simulation data. This computation was done in order to demonstrate that the statistics of the simulated system can be captured by the first terms in the perturbation expansion, but the real significance of our result lies in the physical intuition that can be obtained for the steady-state behavior of systems that reside in this regime.

The zeroth-order expression (\ref{pssgauss}) constitutes a natural extension of the first attempts by Einstein and his contemporaries at understanding macroscopic fluctuations (cf. \cite{Einstein1910}). They interpreted the equilibrium macroscopic distribution to mean that the probability of a fluctuation in a given observable is exponentially suppressed in the ratio of the work that must be done by the rest of the degrees of freedom in the system to produce this fluctuation to the thermal energy scale $k_B T$. Equation (\ref{pssgauss}) simply incorporates the fact that in a driven system, some of the work can be done by the external drive instead of by other internal degrees of freedom. By subtracting off the extra work done by the drive on the way to a given fluctuation (compared to the work done on the way to a fixed reference state), we account for the ``help'' provided by the drive in supplying the work required to reach each state. The first correction term from the perturbation analysis begins to account for the stochasticity in this extra work: even if two possible fluctuations extract the same amount of work on average from the drive, one can be less likely than the other if its distribution of extracted work extends further towards zero, allowing the fluctuation to be reached by paths that receive little help from the drive.

For strictly linear relaxation of a single dissipative current, the mean extra work from the drive can be computed up to an additive constant $C$ in terms of the relaxation time $\tau(E) = 1/A(E)$ using equation (\ref{Wrenpert}) from section \ref{sec:general}: $\Delta \langle W\rangle(J) = V\field J\tau(\field) + C$. For systems close to this regime, the nonlinear response of the steady-state distribution to an increase in the drive can thus be estimated from knowledge of the behavior of the relaxation time. In our sheared colloid case, we can understand the shape of the stress vs. shear rate curve in Figure \ref{fig:worklines} this way. The equilibrium relaxation time $\tau(0)$ is set by the diffusive time scale $ba^2/k_BT$, but in the $\dot{\gamma}\to \infty$ limit, $\tau(\field)$ should be dominated by convective ``stirring'' from the shear flow with time scale $1/\dot{\gamma}$. The shift in the peak of the $\sigma_{xy}^I$ distribution should thus increase linearly with the strength of the conjugate field $\field = -\dot{\gamma}$ near equilibrium, with the expected linear-response coefficient via equation (\ref{linresp}); but the peak should stop shifting once $\tau(\dot{\gamma})$ reaches its asymptotic $1/\dot{\gamma}$ behavior, which cancels out all the $\dot{\gamma}$-dependence. 

As we pointed out in equation (\ref{eq:einstein}) in section \ref{sec:general}, the linear regime also exhibits the remarkable property that the relaxation time $\tau(E)$ and the coefficient $B(E)$ on the noise term in the Langevin equation are related by a generalized Einstein relation or fluctuation-dissipation theorem. A decrease in $\tau(\dot{\gamma})$ must accompanied by an increase in the strength of the noise term in the coarse-grained dynamics, in such a way that the variance of the steady-state distribution remains exactly equal to the variance of the equilibrium distribution. The size of the fluctuations thus remains unchanged in the regime of linear relaxation, even as the dynamics are significantly altered, with a new time scale and (as in the case of this sheared colloid) a new underlying physical mechanism. The first panel of Figure \ref{fig:dist} confirms that this constraint remains applicable even when the linear-response prediction for the mean shear stress is no longer valid.

One could now explore the application of the theory to richer phenomena, such as hydrocluster formation in shear thickening colloids (cf. \cite{Wagner2009,Cheng2011}), where the distribution of some measure of typical cluster size could be computed in terms of the mean work done for trajectories that end at a given value of that parameter. It would also be interesting to investigate whether suspensions of active particles in a quiescent solution can be analyzed in this way, possibly delivering further insight into the ``freezing by heating'' transition that takes place at a critical propulsion rate in both experiment and simulation \cite{Hagan2013,Chaikin2013,Helbing2000}. 

Equation (\ref{pss}) should also be readily generalizable to chemical as opposed to mechanical driving, using the extensions of equation (\ref{crooks1}) to chemical reaction networks mentioned in section \ref{sec:setup} \cite{Qian20052,Qian2013,Gaspard2004}. Once the quantitative relationship between the bulk quantities of interest and the work rate are understood, this generalized result could shed light on the steady-state properties of biologically relevant systems, such as active actin-myosin networks driven by ATP hydrolysis \cite{Martin2014,Ridley2011,Guo2014}. 

\section*{Acknowledgments} 
This research was conducted with Government support under and awarded by DoD, Air Force Office of Scientific Research, National Defense Science and Engineering Graduate (NDSEG) Fellowship, 32 CFR 168a. Benjamin Harpt assisted in debugging the simulation code.

\bibliography{SteadyStateBib,ShearBib,SecondLawBib,library}

\begin{thebibliography}{41}%
\makeatletter
\providecommand \@ifxundefined [1]{%
 \@ifx{#1\undefined}
}%
\providecommand \@ifnum [1]{%
 \ifnum #1\expandafter \@firstoftwo
 \else \expandafter \@secondoftwo
 \fi
}%
\providecommand \@ifx [1]{%
 \ifx #1\expandafter \@firstoftwo
 \else \expandafter \@secondoftwo
 \fi
}%
\providecommand \natexlab [1]{#1}%
\providecommand \enquote  [1]{``#1''}%
\providecommand \bibnamefont  [1]{#1}%
\providecommand \bibfnamefont [1]{#1}%
\providecommand \citenamefont [1]{#1}%
\providecommand \href@noop [0]{\@secondoftwo}%
\providecommand \href [0]{\begingroup \@sanitize@url \@href}%
\providecommand \@href[1]{\@@startlink{#1}\@@href}%
\providecommand \@@href[1]{\endgroup#1\@@endlink}%
\providecommand \@sanitize@url [0]{\catcode `\\12\catcode `\$12\catcode
  `\&12\catcode `\#12\catcode `\^12\catcode `\_12\catcode `\%12\relax}%
\providecommand \@@startlink[1]{}%
\providecommand \@@endlink[0]{}%
\providecommand \url  [0]{\begingroup\@sanitize@url \@url }%
\providecommand \@url [1]{\endgroup\@href {#1}{\urlprefix }}%
\providecommand \urlprefix  [0]{URL }%
\providecommand \Eprint [0]{\href }%
\providecommand \doibase [0]{http://dx.doi.org/}%
\providecommand \selectlanguage [0]{\@gobble}%
\providecommand \bibinfo  [0]{\@secondoftwo}%
\providecommand \bibfield  [0]{\@secondoftwo}%
\providecommand \translation [1]{[#1]}%
\providecommand \BibitemOpen [0]{}%
\providecommand \bibitemStop [0]{}%
\providecommand \bibitemNoStop [0]{.\EOS\space}%
\providecommand \EOS [0]{\spacefactor3000\relax}%
\providecommand \BibitemShut  [1]{\csname bibitem#1\endcsname}%
\let\auto@bib@innerbib\@empty
\bibitem [{\citenamefont {Yamada}\ and\ \citenamefont
  {Kawasaki}(1967)}]{Yamada1967}%
  \BibitemOpen
  \bibfield  {author} {\bibinfo {author} {\bibfnamefont {T.}~\bibnamefont
  {Yamada}}\ and\ \bibinfo {author} {\bibfnamefont {K.}~\bibnamefont
  {Kawasaki}},\ }\href@noop {} {\bibfield  {journal} {\bibinfo  {journal}
  {Prog. Theor. Phys.}\ }\textbf {\bibinfo {volume} {38}},\ \bibinfo {pages}
  {1031} (\bibinfo {year} {1967})}\BibitemShut {NoStop}%
\bibitem [{\citenamefont {Morriss}\ and\ \citenamefont
  {Evans}(1985)}]{Morriss1985}%
  \BibitemOpen
  \bibfield  {author} {\bibinfo {author} {\bibfnamefont {G.~P.}\ \bibnamefont
  {Morriss}}\ and\ \bibinfo {author} {\bibfnamefont {D.~J.}\ \bibnamefont
  {Evans}},\ }\href@noop {} {\bibfield  {journal} {\bibinfo  {journal} {Mol.
  Phys.}\ }\textbf {\bibinfo {volume} {54}},\ \bibinfo {pages} {629} (\bibinfo
  {year} {1985})}\BibitemShut {NoStop}%
\bibitem [{\citenamefont {Morriss}\ and\ \citenamefont
  {Evans}(1988)}]{Morriss1988}%
  \BibitemOpen
  \bibfield  {author} {\bibinfo {author} {\bibfnamefont {G.~P.}\ \bibnamefont
  {Morriss}}\ and\ \bibinfo {author} {\bibfnamefont {D.~J.}\ \bibnamefont
  {Evans}},\ }\href@noop {} {\bibfield  {journal} {\bibinfo  {journal} {Phys.
  Rev. A}\ }\textbf {\bibinfo {volume} {37}},\ \bibinfo {pages} {3605}
  (\bibinfo {year} {1988})}\BibitemShut {NoStop}%
\bibitem [{\citenamefont {Evans}\ and\ \citenamefont
  {Morriss}(2008)}]{Evans2008}%
  \BibitemOpen
  \bibfield  {author} {\bibinfo {author} {\bibfnamefont {D.~J.}\ \bibnamefont
  {Evans}}\ and\ \bibinfo {author} {\bibfnamefont {G.~P.}\ \bibnamefont
  {Morriss}},\ }\href@noop {} {\emph {\bibinfo {title} {Statistical Mechanics
  of Nonequilibrium Liquids}}}\ (\bibinfo  {publisher} {Cambridge University
  Press},\ \bibinfo {address} {Cambridge},\ \bibinfo {year} {2008})\BibitemShut
  {NoStop}%
\bibitem [{\citenamefont {Crooks}(1999)}]{Crooks1999}%
  \BibitemOpen
  \bibfield  {author} {\bibinfo {author} {\bibfnamefont {G.~E.}\ \bibnamefont
  {Crooks}},\ }\href@noop {} {\bibfield  {journal} {\bibinfo  {journal} {Phys.
  Rev. E}\ }\textbf {\bibinfo {volume} {60}},\ \bibinfo {pages} {2721}
  (\bibinfo {year} {1999})}\BibitemShut {NoStop}%
\bibitem [{\citenamefont {McLennan}(1960)}]{McLennan1960}%
  \BibitemOpen
  \bibfield  {author} {\bibinfo {author} {\bibfnamefont {J.~A.}\ \bibnamefont
  {McLennan}},\ }\href {\doibase 10.1063/1.1706081} {\bibfield  {journal}
  {\bibinfo  {journal} {Phys. Fluids}\ }\textbf {\bibinfo {volume} {3}},\
  \bibinfo {pages} {493} (\bibinfo {year} {1960})}\BibitemShut {NoStop}%
\bibitem [{\citenamefont {Lebowitz}\ \emph {et~al.}(1960)\citenamefont
  {Lebowitz}, \citenamefont {Frisch},\ and\ \citenamefont
  {Helfand}}]{Lebowitz1960}%
  \BibitemOpen
  \bibfield  {author} {\bibinfo {author} {\bibfnamefont {J.~L.}\ \bibnamefont
  {Lebowitz}}, \bibinfo {author} {\bibfnamefont {H.~L.}\ \bibnamefont
  {Frisch}}, \ and\ \bibinfo {author} {\bibfnamefont {E.}~\bibnamefont
  {Helfand}},\ }\href {\doibase 10.1063/1.1706037} {\bibfield  {journal}
  {\bibinfo  {journal} {Phys. Fluids}\ }\textbf {\bibinfo {volume} {3}},\
  \bibinfo {pages} {325} (\bibinfo {year} {1960})}\BibitemShut {NoStop}%
\bibitem [{\citenamefont {Komatsu}\ \emph {et~al.}(2009)\citenamefont
  {Komatsu}, \citenamefont {Nakagawa}, \citenamefont {Sasa},\ and\
  \citenamefont {Tasaki}}]{Komatsu2009}%
  \BibitemOpen
  \bibfield  {author} {\bibinfo {author} {\bibfnamefont {T.~S.}\ \bibnamefont
  {Komatsu}}, \bibinfo {author} {\bibfnamefont {N.}~\bibnamefont {Nakagawa}},
  \bibinfo {author} {\bibfnamefont {S.}~\bibnamefont {Sasa}}, \ and\ \bibinfo
  {author} {\bibfnamefont {H.}~\bibnamefont {Tasaki}},\ }\href@noop {}
  {\bibfield  {journal} {\bibinfo  {journal} {J. Stat. Phys.}\ }\textbf
  {\bibinfo {volume} {134}},\ \bibinfo {pages} {401} (\bibinfo {year}
  {2009})}\BibitemShut {NoStop}%
\bibitem [{\citenamefont {Maes}\ and\ \citenamefont
  {Neto\v{c}n\'{y}}(2010)}]{Maes2010}%
  \BibitemOpen
  \bibfield  {author} {\bibinfo {author} {\bibfnamefont {C.}~\bibnamefont
  {Maes}}\ and\ \bibinfo {author} {\bibfnamefont {K.}~\bibnamefont
  {Neto\v{c}n\'{y}}},\ }\href {\doibase 10.1063/1.3274819} {\bibfield
  {journal} {\bibinfo  {journal} {J. Math. Phys.}\ }\textbf {\bibinfo {volume}
  {51}},\ \bibinfo {pages} {015219} (\bibinfo {year} {2010})}\BibitemShut
  {NoStop}%
\bibitem [{\citenamefont {Bertini}\ \emph {et~al.}(2015)\citenamefont
  {Bertini}, \citenamefont {Sole}, \citenamefont {Gabrielli}, \citenamefont
  {Jona-Lasinio},\ and\ \citenamefont {Landim}}]{Bertini2015}%
  \BibitemOpen
  \bibfield  {author} {\bibinfo {author} {\bibfnamefont {L.}~\bibnamefont
  {Bertini}}, \bibinfo {author} {\bibfnamefont {A.~D.}\ \bibnamefont {Sole}},
  \bibinfo {author} {\bibfnamefont {D.}~\bibnamefont {Gabrielli}}, \bibinfo
  {author} {\bibfnamefont {G.}~\bibnamefont {Jona-Lasinio}}, \ and\ \bibinfo
  {author} {\bibfnamefont {C.}~\bibnamefont {Landim}},\ }\href@noop {}
  {\bibfield  {journal} {\bibinfo  {journal} {Rev. Mod. Phys.}\ }\textbf
  {\bibinfo {volume} {47}},\ \bibinfo {pages} {593} (\bibinfo {year}
  {2015})}\BibitemShut {NoStop}%
\bibitem [{\citenamefont {Bertini}\ \emph {et~al.}(2012)\citenamefont
  {Bertini}, \citenamefont {Gabrielli}, \citenamefont {Jona-Lasinio},\ and\
  \citenamefont {Landim}}]{Bertini2012}%
  \BibitemOpen
  \bibfield  {author} {\bibinfo {author} {\bibfnamefont {L.}~\bibnamefont
  {Bertini}}, \bibinfo {author} {\bibfnamefont {D.}~\bibnamefont {Gabrielli}},
  \bibinfo {author} {\bibfnamefont {G.}~\bibnamefont {Jona-Lasinio}}, \ and\
  \bibinfo {author} {\bibfnamefont {C.}~\bibnamefont {Landim}},\ }\href
  {\doibase 10.1007/s10955-012-0624-5} {\bibfield  {journal} {\bibinfo
  {journal} {J. Stat. Phys.}\ }\textbf {\bibinfo {volume} {149}},\ \bibinfo
  {pages} {773} (\bibinfo {year} {2012})}\BibitemShut {NoStop}%
\bibitem [{\citenamefont {Jarzynski}(1997)}]{Jarzynski1997}%
  \BibitemOpen
  \bibfield  {author} {\bibinfo {author} {\bibfnamefont {C.}~\bibnamefont
  {Jarzynski}},\ }\href@noop {} {\bibfield  {journal} {\bibinfo  {journal}
  {Phys. Rev. Lett.}\ }\textbf {\bibinfo {volume} {78}},\ \bibinfo {pages}
  {2690} (\bibinfo {year} {1997})}\BibitemShut {NoStop}%
\bibitem [{\citenamefont {Jarzynski}(2000)}]{Jarzynski2000}%
  \BibitemOpen
  \bibfield  {author} {\bibinfo {author} {\bibfnamefont {C.}~\bibnamefont
  {Jarzynski}},\ }\href@noop {} {\bibfield  {journal} {\bibinfo  {journal} {J.
  Stat. Phys.}\ }\textbf {\bibinfo {volume} {98}},\ \bibinfo {pages} {77}
  (\bibinfo {year} {2000})}\BibitemShut {NoStop}%
\bibitem [{\citenamefont {Qian}(2005)}]{Qian20052}%
  \BibitemOpen
  \bibfield  {author} {\bibinfo {author} {\bibfnamefont {H.}~\bibnamefont
  {Qian}},\ }\href@noop {} {\bibfield  {journal} {\bibinfo  {journal} {J. Phys.
  Chem. B}\ }\textbf {\bibinfo {volume} {109}},\ \bibinfo {pages} {23624}
  (\bibinfo {year} {2005})}\BibitemShut {NoStop}%
\bibitem [{\citenamefont {Ge}\ and\ \citenamefont {Qian}(2013)}]{Qian2013}%
  \BibitemOpen
  \bibfield  {author} {\bibinfo {author} {\bibfnamefont {H.}~\bibnamefont
  {Ge}}\ and\ \bibinfo {author} {\bibfnamefont {H.}~\bibnamefont {Qian}},\
  }\href@noop {} {\bibfield  {journal} {\bibinfo  {journal} {Phys. Rev. E}\
  }\textbf {\bibinfo {volume} {87}},\ \bibinfo {pages} {062125} (\bibinfo
  {year} {2013})}\BibitemShut {NoStop}%
\bibitem [{\citenamefont {Gaspard}(2004)}]{Gaspard2004}%
  \BibitemOpen
  \bibfield  {author} {\bibinfo {author} {\bibfnamefont {P.}~\bibnamefont
  {Gaspard}},\ }\href@noop {} {\bibfield  {journal} {\bibinfo  {journal} {J.
  Chem. Phys.}\ }\textbf {\bibinfo {volume} {120}},\ \bibinfo {pages} {8898}
  (\bibinfo {year} {2004})}\BibitemShut {NoStop}%
\bibitem [{\citenamefont {England}(2013)}]{England2013}%
  \BibitemOpen
  \bibfield  {author} {\bibinfo {author} {\bibfnamefont {J.}~\bibnamefont
  {England}},\ }\href@noop {} {\bibfield  {journal} {\bibinfo  {journal} {J.
  Chem. Phys.}\ }\textbf {\bibinfo {volume} {139}},\ \bibinfo {pages} {121923}
  (\bibinfo {year} {2013})}\BibitemShut {NoStop}%
\bibitem [{\citenamefont {Perunov}\ \emph {et~al.}(2014)\citenamefont
  {Perunov}, \citenamefont {Marsland},\ and\ \citenamefont
  {England}}]{England2014}%
  \BibitemOpen
  \bibfield  {author} {\bibinfo {author} {\bibfnamefont {N.}~\bibnamefont
  {Perunov}}, \bibinfo {author} {\bibfnamefont {R.}~\bibnamefont {Marsland}}, \
  and\ \bibinfo {author} {\bibfnamefont {J.}~\bibnamefont {England}},\
  }\href@noop {} {\bibfield  {journal} {\bibinfo  {journal} {arXiv:1412.1875}\
  } (\bibinfo {year} {2014})},\ \bibinfo {note} {unpublished
  manuscript}\BibitemShut {NoStop}%
\bibitem [{\citenamefont {Gardiner}(2009)}]{Gardiner}%
  \BibitemOpen
  \bibfield  {author} {\bibinfo {author} {\bibfnamefont {C.}~\bibnamefont
  {Gardiner}},\ }\href@noop {} {\emph {\bibinfo {title} {Stochastic
  Methods}}},\ \bibinfo {edition} {4th}\ ed.\ (\bibinfo  {publisher}
  {Springer-Verlag},\ \bibinfo {address} {Berlin},\ \bibinfo {year}
  {2009})\BibitemShut {NoStop}%
\bibitem [{\citenamefont {Fuchs}\ and\ \citenamefont
  {Cates}(2005)}]{Fuchs2005}%
  \BibitemOpen
  \bibfield  {author} {\bibinfo {author} {\bibfnamefont {M.}~\bibnamefont
  {Fuchs}}\ and\ \bibinfo {author} {\bibfnamefont {M.~E.}\ \bibnamefont
  {Cates}},\ }\href {\doibase 10.1088/0953-8984/17/20/003} {\bibfield
  {journal} {\bibinfo  {journal} {J. Phys. Condens. Matter}\ }\textbf {\bibinfo
  {volume} {17}},\ \bibinfo {pages} {S1681} (\bibinfo {year}
  {2005})}\BibitemShut {NoStop}%
\bibitem [{\citenamefont {Harada}(2009)}]{Harada2009}%
  \BibitemOpen
  \bibfield  {author} {\bibinfo {author} {\bibfnamefont {T.}~\bibnamefont
  {Harada}},\ }\href {\doibase 10.1103/PhysRevE.79.030106} {\bibfield
  {journal} {\bibinfo  {journal} {Phys. Rev. E}\ }\textbf {\bibinfo {volume}
  {79}},\ \bibinfo {pages} {030106} (\bibinfo {year} {2009})}\BibitemShut
  {NoStop}%
\bibitem [{\citenamefont {Kr\"{u}ger}\ and\ \citenamefont
  {Fuchs}(2009)}]{Kruger2009}%
  \BibitemOpen
  \bibfield  {author} {\bibinfo {author} {\bibfnamefont {M.}~\bibnamefont
  {Kr\"{u}ger}}\ and\ \bibinfo {author} {\bibfnamefont {M.}~\bibnamefont
  {Fuchs}},\ }\href {\doibase 10.1103/PhysRevLett.102.135701} {\bibfield
  {journal} {\bibinfo  {journal} {Phys. Rev. Lett.}\ }\textbf {\bibinfo
  {volume} {102}},\ \bibinfo {pages} {135701} (\bibinfo {year}
  {2009})}\BibitemShut {NoStop}%
\bibitem [{\citenamefont {Speck}\ and\ \citenamefont
  {Seifert}(2009)}]{Speck2009}%
  \BibitemOpen
  \bibfield  {author} {\bibinfo {author} {\bibfnamefont {T.}~\bibnamefont
  {Speck}}\ and\ \bibinfo {author} {\bibfnamefont {U.}~\bibnamefont
  {Seifert}},\ }\href@noop {} {\bibfield  {journal} {\bibinfo  {journal} {Phys.
  Rev. E}\ }\textbf {\bibinfo {volume} {79}},\ \bibinfo {pages} {040102(R)}
  (\bibinfo {year} {2009})}\BibitemShut {NoStop}%
\bibitem [{\citenamefont {Zhang}\ and\ \citenamefont
  {Szamel}(2011)}]{Zhang2011}%
  \BibitemOpen
  \bibfield  {author} {\bibinfo {author} {\bibfnamefont {M.}~\bibnamefont
  {Zhang}}\ and\ \bibinfo {author} {\bibfnamefont {G.}~\bibnamefont {Szamel}},\
  }\href {\doibase 10.1103/PhysRevE.83.061407} {\bibfield  {journal} {\bibinfo
  {journal} {Phys. Rev. E}\ }\textbf {\bibinfo {volume} {83}},\ \bibinfo
  {pages} {061407} (\bibinfo {year} {2011})}\BibitemShut {NoStop}%
\bibitem [{\citenamefont {Brader}(2010)}]{Brader2010}%
  \BibitemOpen
  \bibfield  {author} {\bibinfo {author} {\bibfnamefont {J.~M.}\ \bibnamefont
  {Brader}},\ }\href@noop {} {\bibfield  {journal} {\bibinfo  {journal} {J.
  Phys. Condens. Matter}\ }\textbf {\bibinfo {volume} {22}},\ \bibinfo {pages}
  {363101} (\bibinfo {year} {2010})}\BibitemShut {NoStop}%
\bibitem [{\citenamefont {Ermak}\ and\ \citenamefont
  {McCammon}(1978)}]{Ermak1978}%
  \BibitemOpen
  \bibfield  {author} {\bibinfo {author} {\bibfnamefont {D.~L.}\ \bibnamefont
  {Ermak}}\ and\ \bibinfo {author} {\bibfnamefont {J.~A.}\ \bibnamefont
  {McCammon}},\ }\href@noop {} {\bibfield  {journal} {\bibinfo  {journal} {J.
  Chem. Phys.}\ }\textbf {\bibinfo {volume} {69}},\ \bibinfo {pages} {1352}
  (\bibinfo {year} {1978})}\BibitemShut {NoStop}%
\bibitem [{\citenamefont {Brady}\ and\ \citenamefont
  {Bossis}(1988)}]{Brady1988}%
  \BibitemOpen
  \bibfield  {author} {\bibinfo {author} {\bibfnamefont {J.~F.}\ \bibnamefont
  {Brady}}\ and\ \bibinfo {author} {\bibfnamefont {G.}~\bibnamefont {Bossis}},\
  }\href@noop {} {\bibfield  {journal} {\bibinfo  {journal} {Annu. Rev. Fluid
  Mech.}\ }\textbf {\bibinfo {volume} {20}},\ \bibinfo {pages} {111} (\bibinfo
  {year} {1988})}\BibitemShut {NoStop}%
\bibitem [{sup()}]{supp2}%
  \BibitemOpen
  \href@noop {} {}\bibinfo {note} {See Supplemental Material at [URL will be
  inserted by publisher] for Python scripts.}\BibitemShut {Stop}%
\bibitem [{\citenamefont {Einstein}(1910)}]{Einstein1910}%
  \BibitemOpen
  \bibfield  {author} {\bibinfo {author} {\bibfnamefont {A.}~\bibnamefont
  {Einstein}},\ }\href@noop {} {\bibfield  {journal} {\bibinfo  {journal} {Ann.
  Phys.}\ }\textbf {\bibinfo {volume} {33}},\ \bibinfo {pages} {1275} (\bibinfo
  {year} {1910})}\BibitemShut {NoStop}%
\bibitem [{\citenamefont {Wagner}\ and\ \citenamefont
  {Brady}(2009)}]{Wagner2009}%
  \BibitemOpen
  \bibfield  {author} {\bibinfo {author} {\bibfnamefont {N.~J.}\ \bibnamefont
  {Wagner}}\ and\ \bibinfo {author} {\bibfnamefont {J.~F.}\ \bibnamefont
  {Brady}},\ }\href@noop {} {\bibfield  {journal} {\bibinfo  {journal} {Phys.
  Today}\ }\textbf {\bibinfo {volume} {62}},\ \bibinfo {pages} {27} (\bibinfo
  {year} {2009})}\BibitemShut {NoStop}%
\bibitem [{\citenamefont {Cheng}\ \emph {et~al.}(2011)\citenamefont {Cheng},
  \citenamefont {McCoy}, \citenamefont {Israelachvili},\ and\ \citenamefont
  {Cohen}}]{Cheng2011}%
  \BibitemOpen
  \bibfield  {author} {\bibinfo {author} {\bibfnamefont {X.}~\bibnamefont
  {Cheng}}, \bibinfo {author} {\bibfnamefont {J.~H.}\ \bibnamefont {McCoy}},
  \bibinfo {author} {\bibfnamefont {J.~N.}\ \bibnamefont {Israelachvili}}, \
  and\ \bibinfo {author} {\bibfnamefont {I.}~\bibnamefont {Cohen}},\
  }\href@noop {} {\bibfield  {journal} {\bibinfo  {journal} {Science}\ }\textbf
  {\bibinfo {volume} {333}},\ \bibinfo {pages} {1276} (\bibinfo {year}
  {2011})}\BibitemShut {NoStop}%
\bibitem [{\citenamefont {Redner}\ \emph {et~al.}(2013)\citenamefont {Redner},
  \citenamefont {Hagan},\ and\ \citenamefont {Baskaran}}]{Hagan2013}%
  \BibitemOpen
  \bibfield  {author} {\bibinfo {author} {\bibfnamefont {G.~S.}\ \bibnamefont
  {Redner}}, \bibinfo {author} {\bibfnamefont {M.~F.}\ \bibnamefont {Hagan}}, \
  and\ \bibinfo {author} {\bibfnamefont {A.}~\bibnamefont {Baskaran}},\
  }\href@noop {} {\bibfield  {journal} {\bibinfo  {journal} {Phys. Rev. Lett.}\
  }\textbf {\bibinfo {volume} {110}},\ \bibinfo {pages} {055701} (\bibinfo
  {year} {2013})}\BibitemShut {NoStop}%
\bibitem [{\citenamefont {Palacci}\ \emph {et~al.}(2013)\citenamefont
  {Palacci}, \citenamefont {Sacanna}, \citenamefont {Steinberg}, \citenamefont
  {Pine},\ and\ \citenamefont {Chaikin}}]{Chaikin2013}%
  \BibitemOpen
  \bibfield  {author} {\bibinfo {author} {\bibfnamefont {J.}~\bibnamefont
  {Palacci}}, \bibinfo {author} {\bibfnamefont {S.}~\bibnamefont {Sacanna}},
  \bibinfo {author} {\bibfnamefont {A.~P.}\ \bibnamefont {Steinberg}}, \bibinfo
  {author} {\bibfnamefont {D.~J.}\ \bibnamefont {Pine}}, \ and\ \bibinfo
  {author} {\bibfnamefont {P.~M.}\ \bibnamefont {Chaikin}},\ }\href@noop {}
  {\bibfield  {journal} {\bibinfo  {journal} {Science}\ }\textbf {\bibinfo
  {volume} {339}},\ \bibinfo {pages} {936} (\bibinfo {year}
  {2013})}\BibitemShut {NoStop}%
\bibitem [{\citenamefont {Helbing}\ \emph {et~al.}(2000)\citenamefont
  {Helbing}, \citenamefont {Farkas},\ and\ \citenamefont
  {Vicsek}}]{Helbing2000}%
  \BibitemOpen
  \bibfield  {author} {\bibinfo {author} {\bibfnamefont {D.}~\bibnamefont
  {Helbing}}, \bibinfo {author} {\bibfnamefont {I.~J.}\ \bibnamefont {Farkas}},
  \ and\ \bibinfo {author} {\bibfnamefont {T.}~\bibnamefont {Vicsek}},\
  }\href@noop {} {\bibfield  {journal} {\bibinfo  {journal} {Phys. Rev. Lett.}\
  }\textbf {\bibinfo {volume} {84}},\ \bibinfo {pages} {1240} (\bibinfo {year}
  {2000})}\BibitemShut {NoStop}%
\bibitem [{\citenamefont {Martin}\ and\ \citenamefont
  {Goldstein}(2014)}]{Martin2014}%
  \BibitemOpen
  \bibfield  {author} {\bibinfo {author} {\bibfnamefont {A.~C.}\ \bibnamefont
  {Martin}}\ and\ \bibinfo {author} {\bibfnamefont {B.}~\bibnamefont
  {Goldstein}},\ }\href@noop {} {\bibfield  {journal} {\bibinfo  {journal}
  {Development}\ }\textbf {\bibinfo {volume} {141}},\ \bibinfo {pages} {1987}
  (\bibinfo {year} {2014})}\BibitemShut {NoStop}%
\bibitem [{\citenamefont {Ridley}(2011)}]{Ridley2011}%
  \BibitemOpen
  \bibfield  {author} {\bibinfo {author} {\bibfnamefont {A.~J.}\ \bibnamefont
  {Ridley}},\ }\href@noop {} {\bibfield  {journal} {\bibinfo  {journal} {Cell}\
  }\textbf {\bibinfo {volume} {145}},\ \bibinfo {pages} {1012} (\bibinfo {year}
  {2011})}\BibitemShut {NoStop}%
\bibitem [{\citenamefont {Guo}\ \emph {et~al.}(2014)\citenamefont {Guo},
  \citenamefont {Ehrlicher}, \citenamefont {Jensen}, \citenamefont {Renz},
  \citenamefont {Moore}, \citenamefont {Goldman}, \citenamefont
  {Lippincott-Schwartz}, \citenamefont {Mackintosh},\ and\ \citenamefont
  {Weitz}}]{Guo2014}%
  \BibitemOpen
  \bibfield  {author} {\bibinfo {author} {\bibfnamefont {M.}~\bibnamefont
  {Guo}}, \bibinfo {author} {\bibfnamefont {A.~J.}\ \bibnamefont {Ehrlicher}},
  \bibinfo {author} {\bibfnamefont {M.~H.}\ \bibnamefont {Jensen}}, \bibinfo
  {author} {\bibfnamefont {M.}~\bibnamefont {Renz}}, \bibinfo {author}
  {\bibfnamefont {J.~R.}\ \bibnamefont {Moore}}, \bibinfo {author}
  {\bibfnamefont {R.~D.}\ \bibnamefont {Goldman}}, \bibinfo {author}
  {\bibfnamefont {J.}~\bibnamefont {Lippincott-Schwartz}}, \bibinfo {author}
  {\bibfnamefont {F.~C.}\ \bibnamefont {Mackintosh}}, \ and\ \bibinfo {author}
  {\bibfnamefont {D.~A.}\ \bibnamefont {Weitz}},\ }\href@noop {} {\bibfield
  {journal} {\bibinfo  {journal} {Cell}\ }\textbf {\bibinfo {volume} {158}},\
  \bibinfo {pages} {822} (\bibinfo {year} {2014})}\BibitemShut {NoStop}%
\bibitem [{\citenamefont {Batchelor}(1970)}]{Batchelor1970}%
  \BibitemOpen
  \bibfield  {author} {\bibinfo {author} {\bibfnamefont {G.~K.}\ \bibnamefont
  {Batchelor}},\ }\href@noop {} {\bibfield  {journal} {\bibinfo  {journal} {J.
  Fluid Mech.}\ }\textbf {\bibinfo {volume} {41}},\ \bibinfo {pages} {545}
  (\bibinfo {year} {1970})}\BibitemShut {NoStop}%
\bibitem [{\citenamefont {Batchelor}(1977)}]{Batchelor1977}%
  \BibitemOpen
  \bibfield  {author} {\bibinfo {author} {\bibfnamefont {G.~K.}\ \bibnamefont
  {Batchelor}},\ }\href@noop {} {\bibfield  {journal} {\bibinfo  {journal} {J.
  Fluid Mech.}\ }\textbf {\bibinfo {volume} {83}},\ \bibinfo {pages} {97}
  (\bibinfo {year} {1977})}\BibitemShut {NoStop}%
\bibitem [{\citenamefont {Griffiths}(1999)}]{Griffiths1999}%
  \BibitemOpen
  \bibfield  {author} {\bibinfo {author} {\bibfnamefont {D.}~\bibnamefont
  {Griffiths}},\ }\href@noop {} {\emph {\bibinfo {title} {Introduction to
  Electrodynamics}}}\ (\bibinfo  {publisher} {Prentice Hall},\ \bibinfo
  {address} {Upper Saddle River, New Jersey},\ \bibinfo {year}
  {1999})\BibitemShut {NoStop}%
\bibitem [{\citenamefont {Lorentz}(1896)}]{Lorentz1906}%
  \BibitemOpen
  \bibfield  {author} {\bibinfo {author} {\bibfnamefont {H.~A.}\ \bibnamefont
  {Lorentz}},\ }\href@noop {} {\bibfield  {journal} {\bibinfo  {journal}
  {Versl. Konigl. Akad. Wetensch. Amst.}\ }\textbf {\bibinfo {volume} {5}},\
  \bibinfo {pages} {168} (\bibinfo {year} {1896})}\BibitemShut {NoStop}%
\end{thebibliography}%

\appendix*
\section{Derivation of Mean Wall Stress (Equations \ref{sigmafull} and \ref{sigmai})}

Formulas for determining the particle contribution to the shear stress of a colloidal suspension have been known for a long time, and received an especially careful treatment by G.K. Batchelor in the 1970's \cite{Batchelor1970,Batchelor1977}. The established literature mainly deals with the \emph{mean} shear stress, either averaged over an infinite ensemble of systems or over an infinitely large system. The statistical uniformity of the system can then be invoked to argue that the mean stress over a typical 2-D slice through the system is equal to the mean stress averaged over the whole system volume. Although the wall is not a \emph{typical} 2-D slice, because the boundary condition modifies the particle distribution, the fact that there is no mean net force on any part of the system when it is in steady state implies that the mean stress on all parallel 2-D slices must be the same. The average over an infinite system volume must therefore also be equal to the average over an infinite wall \cite{Batchelor1970}.

For the purpose of this paper, it is not enough to know the ensemble- or infinite-system-averaged mean. We need to look at the fluctuations about the mean in order to apply our procedure for empirically determining the mean renormalized work and the equilibrium free energy as a function of the shear stress. Therefore we need to go back through the derivation, and examine the \emph{instantaneous} value of the shear stress at the wall in a suspension of a \emph{finite} number of particles.

In this appendix, we prove that the instantaneous shear stress exerted by the fluid on the moving wall of the shear apparatus described in section III, averaged over the moving wall area, is
\begin{align}
\sigma_{xy}^{\rm wall} = \sigma_{xy}^I + \sigma_{xy}^0
\end{align}
where $\sigma_{xy}^I$ is defined by
\begin{align}
\label{sigmadef}
\sigma_{xy}^I \equiv \frac{1}{2V}\sum_{i\neq j} \hat{\mathbf{x}}\cdot\mathbf{F}_{ij} \Delta y_{ij}.
\end{align} 
and $\sigma_{xy}^0$ is independent of the particle positions.

We start by giving some necessary background on the behavior of shear stress in low-Re Newtonian fluids. To make this proof accessible to readers less familiar with hydrodynamics, we then map to a mathematically analogous problem in electrostatics (which turns out to be a homework problem from Griffiths \emph{Electricity and Magnetism} \cite[problem 3.44a]{Griffiths1999}). After presenting the solution to this electrostatics problem, we finally map back to hydrodynamics to obtain our final result.

\subsection{Stress in Newtonian Fluids}
The shear stress $\sigma_{xy}$ is an off-diagonal component of the 3-by-3 stress tensor $\mathbf{\sigma}$. $\mathbf{\sigma}$ is defined at each point in the fluid such that $\hat{\mathbf{n}}\cdot\mathbf{\sigma}$ is the force per unit area exerted from below on a surface element at that location with unit normal vector $\hat{\mathbf{n}}$. By ``from below,'' we mean from the side opposite to the direction of the normal vector. We will focus on the $x$ column $\mathbf{\sigma}\cdot\hat{\mathbf{x}}$ to obtain a vectorial quantity that will be easier to visualize. 

By the definition of the stress tensor above, the $x$-component of the force on a region $\Omega$ of fluid is given by
\begin{align}
\label{fx1}
F_x &= -\int_{\partial\Omega} d\mathbf{A}\cdot \mathbf{\sigma}\cdot\hat{\mathbf{x}}\\
\label{fx2}
& = -\int_{\Omega}dV \nabla \cdot \mathbf{\sigma}\cdot\hat{\mathbf{x}}
\end{align}
where $d\mathbf{A}$ is an infinitesimal area element of the boundary $\partial\Omega$ pointing along the outward normal direction, and $dV$ is an infinitesimal volume element. We add the minus sign because we are computing the force on this surface from the outside. The second line results from the divergence theorem. Since this holds for every possible region $\Omega$, we conclude that the integrand of equation (\ref{fx2}) is equal to minus the $x$ component of force per unit volume $f_x$ exerted by the surrounding fluid on an infinitesimal volume element:
\begin{align}
\nabla \cdot \mathbf{\sigma}\cdot\hat{\mathbf{x}} = -f_x.
\end{align}
Finally, we must invoke the assumption that the solvent in which the particles are suspended is a Newtonian fluid, which implies
\begin{align}
\mathbf{\sigma}\cdot\hat{\mathbf{x}} = - \eta_0 \nabla u_x
\end{align}
where $u_x$ is the $x$-component of the fluid velocity field, and $\eta_0$ is the (constant) viscosity of the solvent. Combining this with the previous equation gives us the set of equations
\begin{align}
\eta_0\nabla^2 u_x& =  f_x \label{sigma1}\\
\mathbf{\sigma}\cdot\hat{\mathbf{x}}& = - \eta_0 \nabla u_x\label{sigma2}
\end{align}
that together fully determine $\mathbf{\sigma}\cdot\hat{\mathbf{x}}$ for a given set of boundary conditions.

\subsection{Mapping to Electrostatics}
Equations (\ref{sigma1}) and (\ref{sigma2}) suggest a mapping to electrostatics. $\eta_0 u_x$ is the analog to the electric potential $\phi$, $\mathbf{\sigma}\cdot\hat{\mathbf{x}}$ is the analog to the electric field $\mathbf{E}$, and $-f_x$ is the analogue to the charge density $\rho$. With these mappings, the mathematics of the problem are identical to electrostatics, and we can do everything in terms of $\mathbf{E}$, $\phi$ and $\rho$ until we map back at the end.

The only remaining piece of setup is to map the boundary conditions and the ``charge distribution.'' The non-slip boundary condition requires that every part of the fluid in contact with a non-rotating rigid surface must share the same velocity. Since the electric potential $\phi$ is the analog of the $x$-component of velocity, this implies that non-rotating surfaces behave like perfect conductors - they are always equipotentials. In particular, the constraint that the bottom wall is fixed and the top wall moves at constant velocity $v$ implies that the walls of the shear cell become parallel conducting plates separated by a distance $d$, with fixed electric potential difference $\Delta \phi$. The problem of determining the total force on the walls is thus equivalent to determining the induced charge on these conducting plates.

The particles, however, are allowed to rotate. Their boundary conditions are therefore more complicated, involving the other columns of the stress tensor. Specifically, we have
\begin{align}
  \mathbf{u} = \Omega\times \mathbf{r}_{\perp}+\mathbf{u}_{\rm cm}
\end{align}
for all points on the surface of the sphere, where $\mathbf{r}_{\perp}$ is the vector pointing from the center of the sphere to the surface point, projected onto a plane perpendicular to the angular velocity vector $\Omega$. $\Omega$ and the center-of-mass velocity $\mathbf{u}_{\rm cm}$ are free parameters that must be adjusted so as to be consistent with equations (\ref{sigma1}) and (\ref{sigma2}). The resulting restriction on $\mathbf{\sigma}$ is 
\begin{align}
  \mathbf{\sigma} = -\eta_0 \nabla\left( \Omega\times \mathbf{r}_{\perp}+\mathbf{u}_{\rm cm}\right).
\end{align}

To determine the charge distribution, we use our assumption of low Re to require the total force on any volume element to vanish. In the electrostatic analogy, this implies that the solvent is uncharged, and all charge must reside at the walls or on the particles. The interparticle repulsion exerts force on each particle that must be canceled by the friction of the fluid in order to satisfy the requirement of zero total force. This implies that the total ``charge'' on each particle must be $q_i = \sum_{j\neq i}\mathbf{F}_{ji}\cdot\hat{\mathbf{x}}$, where $\mathbf{F}_{ji}$ is the force exerted on particle $i$ by particle $j$. The distribution of this total charge over the surface of each sphere is not fixed in advance, however, and must be determined by solving equations (\ref{sigma1}) and (\ref{sigma2}) (along with the corresponding equations for the other components of the stress tensor) with the boundary conditions just described. The decision to ``ignore hydrodynamic interactions'' mentioned in the main text allows us to greatly simplify the problem of determining these distributions, by solving the equations for each particle individually, with boundary condition $\mathbf{E} \to -(\Delta \phi/d)\hat{\mathbf{y}}$ far from the sphere. This approximation ignores the effect of the other particles and of the induced wall charge on the charge distribution over each sphere. The solutions obtained under this approximation are independent of the particle positions, which will be important later on.

\subsection{Obtaining the Induced Charge on the Conducting Plate}
Our problem is thus reduced to determining the induced charge on a pair of conducting parallel plates at fixed electric potential due to a given charge distribution inside. 

We start by splitting the charge on the plates into two parts, following the strategy of Batchelor in his treatment of the effect of particle interactions on mean shear stress \cite{Batchelor1977}. The derivation will resemble Batchelor's in many ways, despite the electrostatic language, but adds the a new element by considering the wall stress due to a given \emph{instantaneous} configuration of particles as opposed to an ensemble average of all possible configurations.

The first part of the charge is the part required to maintain the electric potential difference $\Delta \phi$ in the absence of any additional charges between the plates: $Q_0 = A \Delta \phi/d$ on the top and $-Q_0$ on the bottom. To find the remaining charge, we can solve for the case where the two plates are grounded. When we add up the two charge distributions, the resulting field is guaranteed to produce the desired constant electric potential difference. The case of grounded plates is problem 3.44a in Griffiths, as mentioned above, and we will follow his method to solve it \cite{Griffiths1999}.

Griffiths starts by having the student derive a relation known as Green's Reciprocity Theorem. (This theorem is closely related to a result due to Lorentz in hydrodynamics, which Batchelor employs in his analysis \cite{Lorentz1906}.) Consider two distinct charge distributions $\rho_1(\mathbf{r})$ and $\rho_2(\mathbf{r})$, which produce electric fields $\mathbf{E}_1(\mathbf{r})$ and $\mathbf{E}_2(\mathbf{r})$, with electric potentials $\phi_1(\mathbf{r})$ and $\phi_2(\mathbf{r})$. Now use the Maxwell Equation $\nabla \cdot \mathbf{E} = \rho $ and the definition of electric potential $\mathbf{E} = -\nabla \phi$ to obtain
\begin{align}
  \int dV \mathbf{E}_1\cdot\mathbf{E}_2 & = -\int dV \nabla \phi_1 \cdot \mathbf{E}_2  = \int dV\phi_1\nabla \cdot \mathbf{E}_2 \nonumber\\
&= \int dV\phi_1 \rho_2\\
&= -\int dV \nabla \phi_2 \cdot \mathbf{E}_1  = \int dV \phi_2\nabla \cdot \mathbf{E}_1 \nonumber\\
&= \int dV \phi_2 \rho_1
\end{align}
where we are integrating over all space, and have used integration by parts to switch the $\nabla$ from $\phi$ to $\mathbf{E}$.

We thus obtain Green's Reciprocity Theorem:
\begin{align}
  \int dV\phi_1\rho_2 = \int dV \phi_2\rho_1.
\end{align}
Now we use this relation to compute the induced charge on our plates. We will start by computing the induced charge due to a point charge $q$ at location $\mathbf{r}=(x,y,z)$. We will work in coordinates where the bottom plate is at $y=0$ and the top is at $y = d$. 

To apply the Reciprocity Theorem, we choose for $\rho_1$ the actual charge distribution we are analyzing, with the point charge between the grounded parallel plates. We define $Q_+$ as the total induced charge on the top plate and $Q_-$ as the total induced charge on the bottom plate. For $\rho_2$, we choose a charge distribution with conducting plates in the same locations, but with the top plate fixed at electric potential $\phi_0$ above the bottom one, and with no charge in the space between them. The LHS of the Reciprocity Theorem vanishes, because $\phi_1 =0$ whenever $\rho_2$ is nonzero. The RHS has a contribution from the charge distribution on the top plate, and a contribution from the particle. If the plates are infinite, then the potential a distance $y$ above the bottom plate in scenario 2 is exactly $(y/d)\phi_0$. This will still be a good approximation in a finite system for charges that are not too close to the edges of the system, which will be true for the charges on the vast majority of the spheres when the number of spheres is large. Thus we obtain: 
\begin{align}
0 = \phi_0 Q_+ + \phi_0\frac{y}{d}q.
\end{align}
Solving for $Q_+$, we find
\begin{align}
Q_+ = -\frac{y}{d}q.
\end{align}
Now we again use the linearity of our equations to obtain the total induced charge by summing up the contributions from all the infinitesimal charge elements in the distribution. A convenient way to perform this sum is to split up the charge distribution on each sphere into two parts: a spatially uniform part equal to the mean surface charge on the sphere, and spatially varying part that integrates to zero over each sphere surface. 

\paragraph{Contribution of Variations about the Mean} We start by computing the contribution of the second part of the charge distribution. Since this part of the charge sums to zero on each sphere, every positive charge $\delta q$ has a corresponding negative charge $-\delta q$ somewhere else on the sphere. The net induced charge from each such pair is 
\begin{align}
\delta Q_+ = \frac{\delta q}{d}(y_- - y_+)
\end{align}
where $y_-$ and $y_+$ are the coordinates of the $+\delta q$ and $-\delta q$ charges, respectively. Now recall that by ignoring hydrodynamic interactions, we can solve for the charge distribution over each sphere without knowing its position relative to the plates or the other particles. Furthermore, the linearity of the governing equations implies that the variations about the mean charge density are independent of the size of the mean. This implies that the $y$-distance $y_--y_+$ between any pair of charges on a single sphere is independent of the spatial configuration of the particles and of the total charge $q_i$ of the particle in question. 

Summing over all pairs of charges from all the spheres in the sample, we define the quantity
\begin{align}
Q_H = \sum \delta Q_+
\end{align}
as the total induced charge due to the variations about the mean charge on the surface of the spheres. This quantity is independent of the particle positions, and just adds a constant offset to the total charge. The $H$ subscript stands for ``hydrodynamic,'' because this contribution comes purely from the friction of the flow field around each particle.

\paragraph{Contribution of the Mean Charge} To complete our calculation, we must compute the charge induced on the plate by a given configuration of uniformly charged spheres. Since the field of a uniformly charged sphere is equivalent to the field of a point charge (for points outside the surface of the sphere), we can simply evaluate the point charge solution derived above for every particle, and add them all up. We thus find
\begin{align}
Q_I = -\sum_i\frac{y_i}{d}q_i.
\end{align}
Combining the above results, we find that the total induced charge on the top plate is $Q =  Q_I+Q_0 + Q_H $, with $Q_I$ the only term that depends on the particle positions.

\subsection{Mapping Back to Hydrodynamics}
We can now map back into the original variables (recalling that charge is equivalent to minus the force exerted by the fluid) in order to obtain the total force exerted by the fluid on the moving wall of the shear apparatus:
\begin{align}
F_{\rm wall} &= \sum_i \frac{y_i}{d} \left(\sum_{j\neq i}\hat{\mathbf{x}}\cdot\mathbf{F}_{ji}\right)+F_0 + F_H.
\end{align}
We can simplify this expression by using the fact that $\mathbf{F}_{ji} = - \mathbf{F}_{ij}$:
\begin{align}
F_{\rm wall}= \frac{1}{2d}\sum_{i\neq j}\hat{\mathbf{x}}\cdot \mathbf{F}_{ij}\Delta y_{ij} + F_0 + F_H.
\end{align}
Finally, we can divide through by the area $A$ of the wall to obtain the mean shear stress exerted on the wall by the fluid:
\begin{align}
\sigma_{xy}^{\rm wall}&= \sigma_{xy}^I + \sigma_{xy}^0 + \sigma_{xy}^H
\end{align}
where 
\begin{align}
\sigma_{xy}^I = \frac{1}{2V}\sum_{i\neq j}\hat{\mathbf{x}}\cdot \mathbf{F}_{ij}\Delta y_{ij}
\end{align}
and the other two terms are independent of the particle positions. For notational simplicity, we combine them into one term in the main text, which we call $\sigma_{xy}^0$.

\end{document}